\documentclass[aps,pra,twocolumn,superscriptaddress]{revtex4-2}

\usepackage[version=3]{mhchem} % Formula subscripts using \ce{}
\usepackage{siunitx}
\usepackage{xcolor}
\usepackage{color,soul}
\usepackage{graphicx}
\usepackage{mciteplus}
\usepackage{url}
\mciteErrorOnUnknownfalse
\begin{document}

%%%%%%%%%%%%%%%%%%%%%%%%%%%%%%%%%%%%%%%%%%%%%%%%%%%%%%%%%%%%%%%%%%%%%
%% The document title should be given as usual. Some journals require
%% a running title from the author: this should be supplied as an
%% optional argument to \title.
%%%%%%%%%%%%%%%%%%%%%%%%%%%%%%%%%%%%%%%%%%%%%%%%%%%%%%%%%%%%%%%%%%%%%
\title{Nano-Ironing van der Waals Heterostructures Towards Electrically Controlled Quantum Dots}
%%%%%%%%%%%%%%%%%%%%%%%%%%%%%%%%%%%%%%%%%%%%%%%%%%%%%%%%%%%%%%%%%%%%%
%% Some journals require a list of abbreviations or keywords to be
%% supplied. These should be set up here, and will be printed after
%% the title and author information, if needed.
%%%%%%%%%%%%%%%%%%%%%%%%%%%%%%%%%%%%%%%%%%%%%%%%%%%%%%%%%%%%%%%%%%%%%

\author{Teymour Talha-Dean}
    \thanks{These two authors contributed equally}
    \affiliation{Institute of Materials Research and Engineering (IMRE), Agency for Science, Technology and Research (A*STAR), 2 Fusionopolis Way, Innovis $\#$08-03, Singapore 138634, Republic of Singapore}
    \affiliation{Department of Physics and Astronomy, Queen Mary University of London, London, E1 4NS, United Kingdom}

\author{Yaoju Tarn}
    \thanks{These two authors contributed equally}
    \affiliation{Institute of Materials Research and Engineering (IMRE), Agency for Science, Technology and Research (A*STAR), 2 Fusionopolis Way, Innovis $\#$08-03, Singapore 138634, Republic of Singapore}

\author{Subhrajit Mukherjee}
    \affiliation{Institute of Materials Research and Engineering (IMRE), Agency for Science, Technology and Research (A*STAR), 2 Fusionopolis Way, Innovis $\#$08-03, Singapore 138634, Republic of Singapore}

\author{John Wellington John}
    \affiliation{Institute of Materials Research and Engineering (IMRE), Agency for Science, Technology and Research (A*STAR), 2 Fusionopolis Way, Innovis $\#$08-03, Singapore 138634, Republic of Singapore}

\author{Ding Huang}
    \affiliation{Institute of Materials Research and Engineering (IMRE), Agency for Science, Technology and Research (A*STAR), 2 Fusionopolis Way, Innovis $\#$08-03, Singapore 138634, Republic of Singapore}

\author{Ivan A. Verzhbitskiy}
    \affiliation{Institute of Materials Research and Engineering (IMRE), Agency for Science, Technology and Research (A*STAR), 2 Fusionopolis Way, Innovis $\#$08-03, Singapore 138634, Republic of Singapore}

\author{Dasari Venkatakrishnarao}
    \affiliation{Institute of Materials Research and Engineering (IMRE), Agency for Science, Technology and Research (A*STAR), 2 Fusionopolis Way, Innovis $\#$08-03, Singapore 138634, Republic of Singapore}

\author{Sarthak Das}
    \affiliation{Institute of Materials Research and Engineering (IMRE), Agency for Science, Technology and Research (A*STAR), 2 Fusionopolis Way, Innovis $\#$08-03, Singapore 138634, Republic of Singapore}

\author{Rainer Lee}
    \affiliation{Institute of Materials Research and Engineering (IMRE), Agency for Science, Technology and Research (A*STAR), 2 Fusionopolis Way, Innovis $\#$08-03, Singapore 138634, Republic of Singapore}

\author{Abhishek Mishra}
    \affiliation{Institute of Materials Research and Engineering (IMRE), Agency for Science, Technology and Research (A*STAR), 2 Fusionopolis Way, Innovis $\#$08-03, Singapore 138634, Republic of Singapore}

\author{Shuhua Wang}
    \affiliation{Science, Mathematics and Technology, Singapore University of Technology and Design, 8 Somapah Road, 487372, Singapore}

\author{Yee Sin Ang}
    \affiliation{Science, Mathematics and Technology, Singapore University of Technology and Design, 8 Somapah Road, 487372, Singapore}

\author{Kuan Eng Johnson Goh}
    \email{\url{kejgoh@yahoo.com}}
        \affiliation{Institute of Materials Research and Engineering (IMRE), Agency for Science, Technology and Research (A*STAR), 2 Fusionopolis Way, Innovis $\#$08-03, Singapore 138634, Republic of Singapore}
        \affiliation{Department of Physics, National University of Singapore, 2 Science Drive 3, 117551, Singapore}
        \affiliation{Division of Physics and Applied Physics, School of Physical and Mathematical Sciences, Nanyang Technological University, 50 Nanyang Avenue 639798, Singapore}

\author{Chit Siong Lau}
    \email{\url{aaron_lau@imre.a-star.edu.sg}}
        \affiliation{Institute of Materials Research and Engineering (IMRE), Agency for Science, Technology and Research (A*STAR), 2 Fusionopolis Way, Innovis $\#$08-03, Singapore 138634, Republic of Singapore}

\begin{abstract}
Assembling two-dimensional van der Waals layered materials into heterostructures is an exciting development that sparked the discovery of rich correlated electronic phenomena and offers possibilities for designer device applications. However, resist residue from fabrication processes is a major limitation. Resulting disordered interfaces degrade device performance and mask underlying transport physics. Conventional cleaning processes are inefficient and can cause material and device damage. Here, we show that thermal scanning probe based cleaning can effectively eliminate resist residue to recover pristine material surfaces. Our technique is compatible at both the material- and device-level, and we demonstrate the significant improvement in the electrical performance of 2D WS$_2$ transistors. We also demonstrate the cleaning of van der Waals heterostructures to achieve interfaces with low disorder. This enables the electrical formation and control of quantum dots that can be tuned from macroscopic current flow to the single-electron tunnelling regime. Such material processing advances are crucial for constructing high-quality vdW heterostructures that are important platforms for fundamental studies and building blocks for quantum and nano-electronics applications. 

\end{abstract}

%title - 'Nano-Ironing: Toward clean two-dimensional materials and electrically controlled quantum dots'

\maketitle

\section{Introduction}

The successful isolation of graphene with mechanical exfoliation was a profound moment.\cite{Novoselov2004} It led to an explosive growth in the field of two-dimensional (2D) materials. A large and growing `2D library' is now available, with thousands of 2D insulating, semiconducting, and metallic materials offering an extensive array of properties. Subsequent development of the dry transfer technique enabled the stacking of individual 2D layers into van der Waals (vdW) heterostructures.\cite{Haigh2012a,Geim2013a,Pizzocchero2016} Akin to the assembly of simple Lego blocks into complex structures, vdW heterostructures often display properties distinct from each individual layer. This offers exciting avenues for designer devices with tailored functionalities. Beyond applications, vdW heterostructures are ideal platforms for fundamental studies. Strong electronic correlations across layers can lead to emergent physics and quantum effects including unconventional superconductivity,\cite{Cao2018a} ferromagnetism,\cite{Sharpe2019} ferroelectricity,\cite{Zheng2020} and quantized anomalous Hall effect.\cite{Serlin2020} 

Central to these ideas is the need for pristine vdW surfaces/interfaces, but a key problem is resist residue from lithography. Such residue can form films or clusters $\sim$1-10 nm thick, which is significant considering the atomic thinness of 2D materials ($<1$ nm). Several techniques exist for cleaning the resist residue. Standard solvent cleaning is typically insufficient and can introduce unwanted chemical doping.\cite{Dan2009} It is usually supplemented by other techniques such as thermal annealing,\cite{Cheng2011} current annealing,\cite{Bolotin2008} and nano-probe tip based mechanical cleaning\cite{Lindvall2012a,Rosenberger2018,Kim2019c,Palai2023}. The effectiveness of resist removal scales with temperature for thermal annealing, but the device thermal budget is an important consideration as contacts, substrates, and dielectrics may not be compatible with high temperatures.\cite{Liu2021} Current annealing allows local channel heating but can lead to material damage and device breakdown unless complex feedback systems\cite{Lau2014} are introduced. Mechanical cleaning with nano-probe tips can restore atomically clean surfaces but may require large local forces, especially for larger sized residue. These large forces lead to unintended defects and cause ruptures and film tears.\cite{Rosenberger2018,Kim2019c,Palai2023} 

To address the need for a damage-free method of resist removal for atomically clean and flat vdW  material surfaces, we develop a technique we call `nano-ironing'. The nano-ironing concept combines the advantages of annealing with mechanical tip based cleaning by using a heated nano probe tip. We show that nano-ironing is substantially more effective at removing resist residue compared to conventional mechanical nano probe tip based cleaning (`nano-brooming') at similar force levels. By restoring surface cleanliness, WS$_2$ field effect transistors (FET) display significant improvements in carrier mobility and drain current, and reduction in unwanted charge doping. Finally, we use nano-ironing to prepare vdW heterostructures with low disorder. We show quantum transport measurements where we demonstrate electrical confinement and control of carriers to observe a transition from macroscopic current flow towards single-electron transport. 

\section{Nano-Ironing: Thermal Scanning Probe-Based Cleaning}

\begin{figure*}
    \centering
    \includegraphics[width=15cm]{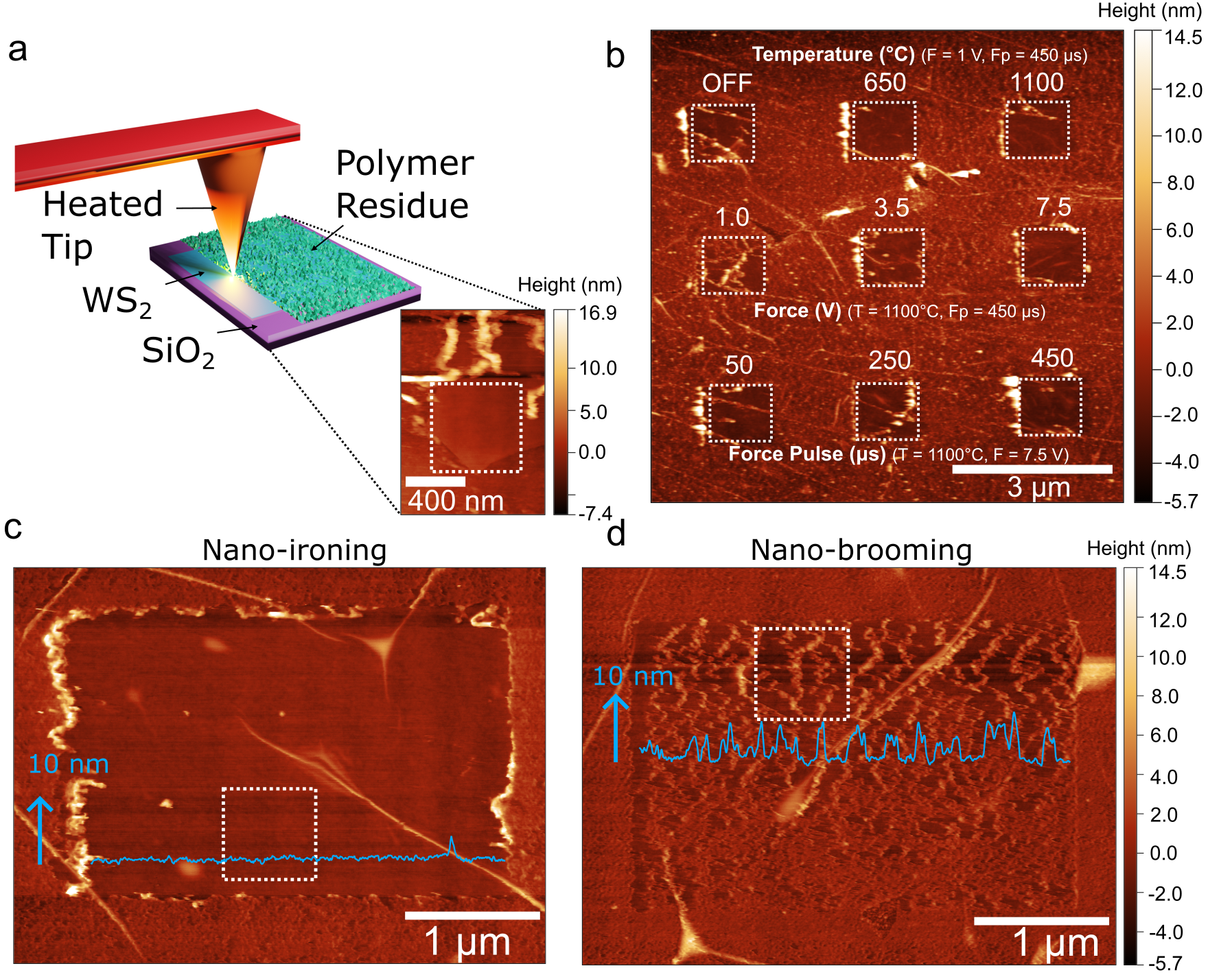}
    \caption{\textbf{Nano-ironing: heated nano-probe tip based cleaning of 2D material surfaces} (a) Schematic of the nano-ironing process. The inset shows an atomic-force microscopy (AFM) image after nano-ironing analogous to the cartoon schematic. The white box represents the atomically smooth nano-ironed flake area. Surrounding regions are covered with polymeric resist residue. (b) Demonstration of tuning parameters that can influence the degree of resist removal. Writer temperature $T$ varies from room temperature to 1100\textsuperscript{$\circ$}C, corresponding to an estimated range of tip contact temperatures of 30 to 250\textsuperscript{$\circ$}C.\cite{Zheng2019a} Applied forces, $F$, are expressed in voltage applied to the piezo tip. From these voltages, we can estimate the forces in newtons (Supporting Information). Force pulse, $F_\mathrm{p}$ is the duration for which the heated tip is in contact with each pixel on the surface. (c,d) Comparison of nano-ironed (with heat) and nano-broomed (without heat) areas at comparable $F$ and $F_\mathrm{P}$ showing a substantially cleaner surface after nano-ironing. The AFM images in c and d share the same height scale bar. Respective line cuts are overlayed in blue, with the height scale indicated by the blue arrow. Root-mean-square (RMS) roughness values are calculated from the areas enclosed by the dashed squares, with 300$\pm$40 pm and 1.20$\pm$0.3 nm for nano-ironing and nano-brooming respectively.}
    \label{fig:1}
\end{figure*}

Our nano-ironing technique is based on a commercial thermal scanning probe lithography (t-SPL) system housed in an inert glovebox with typical O$_2$ and H$_2$O levels below 5 ppm.\cite{Albisetti2022,Zheng2019a} Such inert environment mitigates possible oxidation of 2D material films during nano-ironing. The t-SPL's heated nanoprobe tip comes into contact with a vdW material surface and rasters to thermo-mechanically clean polymeric resist (Figure~\ref{fig:1}a). Precise alignment and controlled cleaning are possible with tunability of tip temperature, applied force, and duration of contact. (Figure~\ref{fig:1}b). We find that increasing the temperature can improve resist removal efficiency, while higher forces are useful for particularly stubborn or larger sized particulates. A higher force pulse, i.e., the duration of heated contact, prolongs heat transfer for better cleaning.

The key advantage of nano-ironing over nano-brooming lies in the extra degree of freedom allotted by temperature control. This achieves a substantially cleaner surface at comparable forces (Figure~\ref{fig:1}c, d) with almost an order of magnitude improvement in root-mean-square (RMS) roughness. We recover a surface with RMS roughness of 300$\pm 40$ pm (1.20$\pm 0.3$ nm) after nano-ironing (nano-brooming). Importantly, nano-ironing does not lead to any observable chemical and mechanical damage as shown via photoluminescence and Raman spectroscopy measurements (Supporting Information Figure S3). Detailed discussion on force and contact temperature estimations, and choice of t-SPL parameters for optimal cleanliness is available in the Supporting Information. 

\section{Electrical Characterization}

\begin{figure*}
    \centering
    \includegraphics[width=15cm]{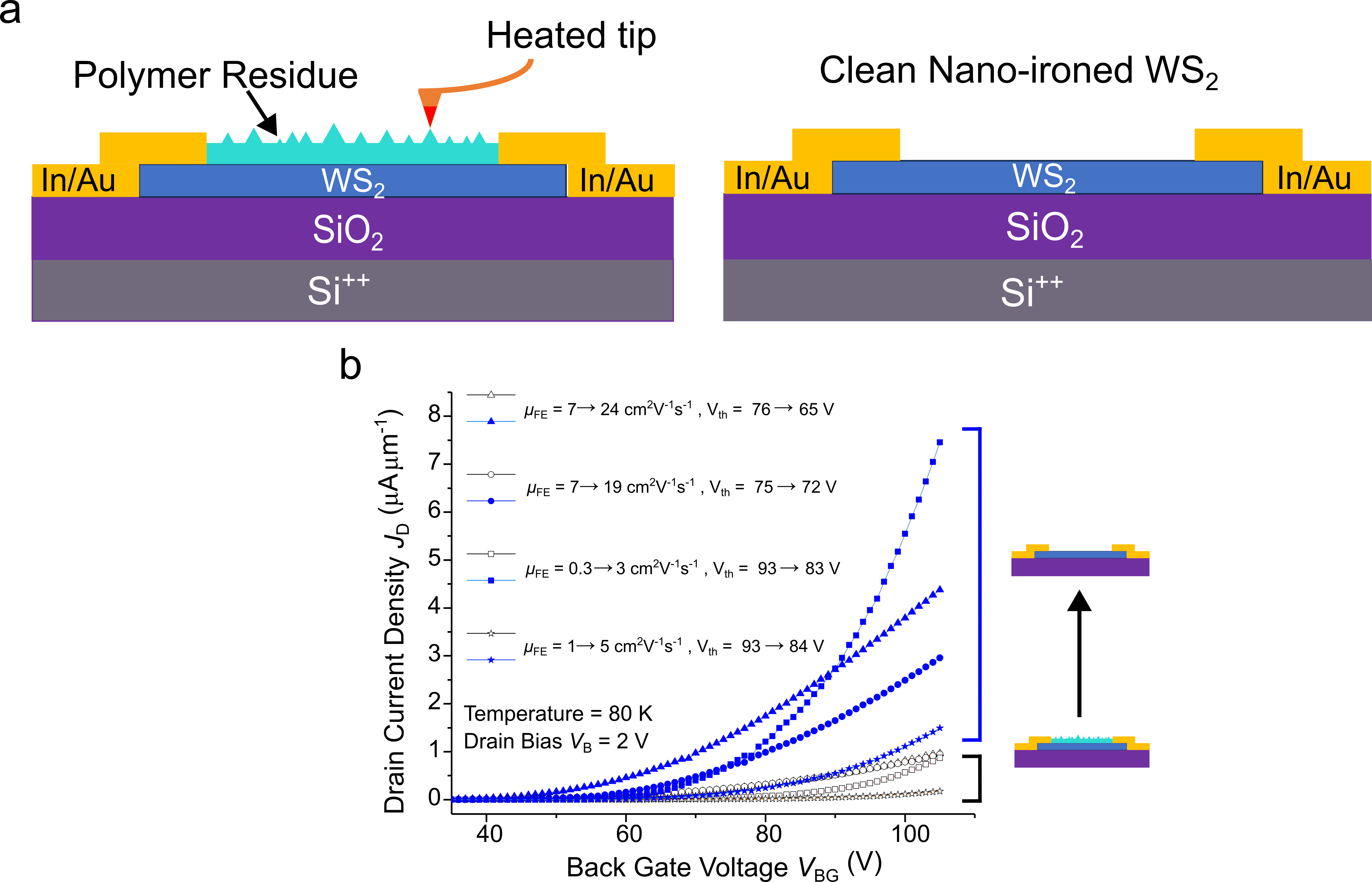}
    \caption{\textbf{Electrical transport improvements from nano-ironing}. (a) Schematics of a back-gated 2D WS$_2$ field effect transistor subjected to the nano-ironing process. (b) Transfer curves measured at 80 K for 4 separate devices before (black, unfilled) and after (blue, filled) nano-ironing. Carrier field-effect mobilities $\mu_\mathrm{FE}$ and threshold voltages $V_\mathrm{th}$ determined from the linear regimes show significant changes from nano-ironing. We observe between $\sim$2 to 10-fold improvements in $\mu_\mathrm{FE}$ and negative shifts of $\sim$3 to 11 V in $V_\mathrm{th}$.}
    \label{fig:2}
\end{figure*}

Polymeric resist residue can degrade carrier transport in 2D FETs through unwanted charge doping and increased roughness and impurity scattering.\cite{Pirkle2011,Suk2013} A useful way to estimate the effects of scattering mechanisms on carrier mobility $\mu_\mathrm{FE}$ in 2D FETs is Matthiesen’s rule $\frac{1}{\mu} = \sum\limits_{i} \frac{1}{\mu_\mathrm{i}}$, where $\mu_\mathrm{i}$ is the contribution from the $i$th scattering mechanism. The main scattering mechanisms are impurities, interface roughness, and phonons.\cite{Lau2021,Kieczka2023} At room temperature, $\mu$ is typically limited by electron-phonon scattering which can mask potential improvements from nano-ironing. Therefore, we investigate the low temperature performance of 2D WS$_2$ FETs before and after nano-ironing. We show that nano-ironing can indeed lead to increased current densities, increased carrier mobilities, and reduced unwanted charge-doping. 

Our device consists of a chemical vapour deposition (CVD) grown monolayer WS$_\mathrm{2}$ semiconducting channel with indium-gold (In/Au) alloy contacts on a heavily doped SiO$_\mathrm{2}$/Si substrate (Figure 2a). The SiO$_\mathrm{2}$/Si substrate also functions as a back gate electrode. Transfer curves of a 2D WS$_2$ FET measured at 80 K before (black, unfilled) and after (blue, filled) nano-ironing show increases in drain current density $J_\mathrm{D}$ and field effect mobility $\mu_\mathrm{FE}$ of $\sim$2 to 10- fold across four devices. This highlights the improvements from reduced impurity and roughness scattering (Figure 2b). We also observe negative shifts in threshold voltage $V_\mathrm{th}$ of $\sim$3 to 11 V post nano-ironing, consistent with the expected reduction in $p$-type charge doping after eliminating resist residue.\cite{Pirkle2011}

\section{Quantum Dot Control}

\begin{figure*}
    \centering
    \includegraphics[width=15cm]{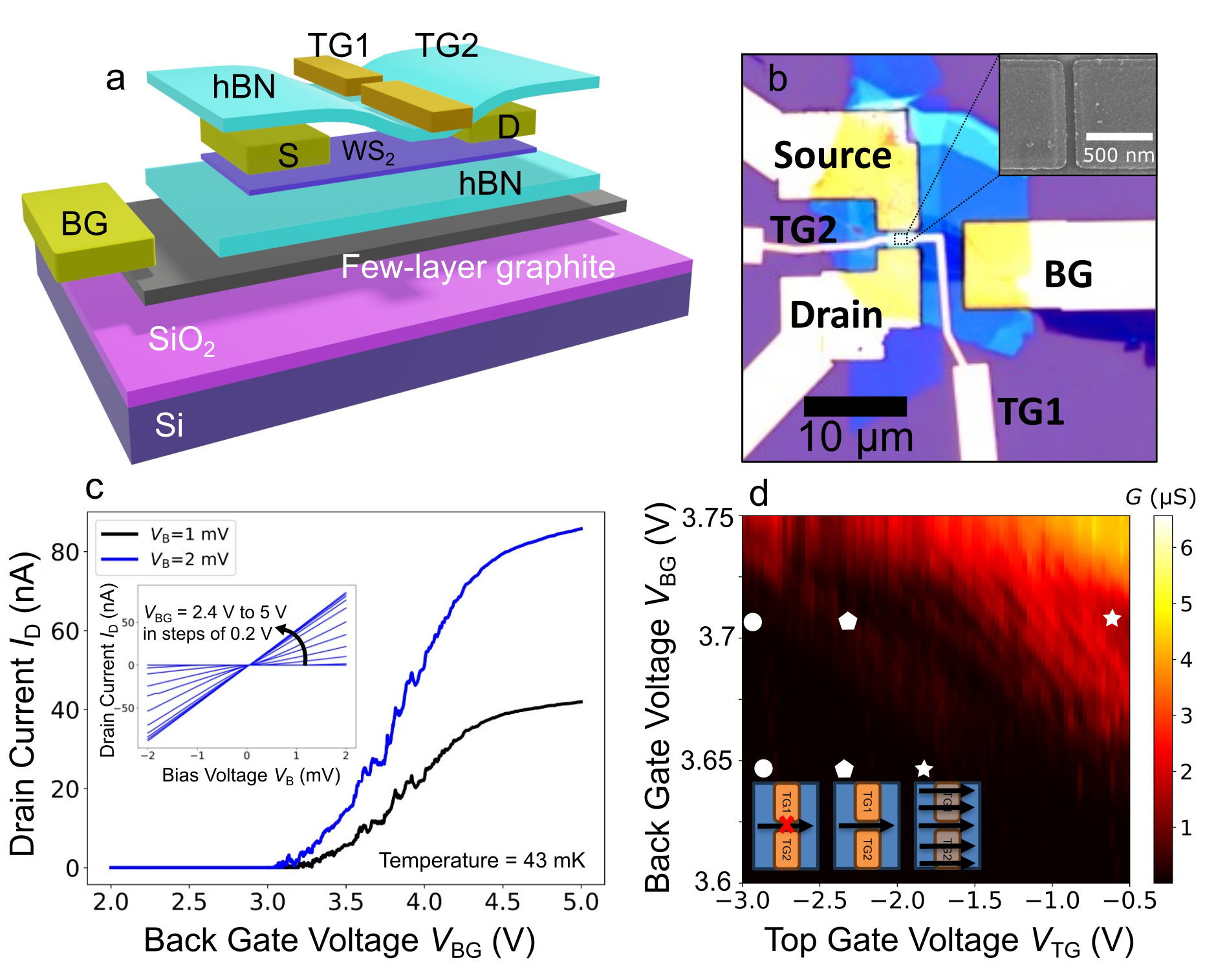}
    \caption{\textbf{Nano-ironing two-dimensional van der Waals heterostructure}. (a) Schematic illustration of our device consisting of a few-layer graphite back gate (BG), hBN/WS$_2$/hBN heterostructure with source-drain contacts (S/D), and split top gate electrodes (TG1/TG2). (b) Optical image of the device with labelled electrodes. Scale bar 10 $\mu$m.  The inset shows a scanning electron micrograph of the split top gates separated by a gap of $\sim$70 nm. The few-layer graphene back gate (BG) modulates the overall carrier density in the WS$_2$ channel and is separated by a 10 nm thick hBN dielectric. Another 16 nm thick top hBN dielectric layer electrically insulates the WS$_2$ channel from the split top gates used for local modulation of the electrostatic potential in the WS$_2$ channel. Transport measurements of our van der Waals heterostructure at a temperature of 43 mK. (c) Transfer curves showing typical $n$-type transistor characteristics. The inset shows the output curves where linear ohmic behaviour highlights the high quality In/Au contacts with low Schottky barrier heights. (d) Device conductance $G$ as a function of the back gate voltage $V_\mathrm{BG}$ and a common top gate voltage $V_\mathrm{TG}$ applied to both split top gates. Schematic illustrations of the transport regimes indicated by the white circle, pentagon, and star symbols are shown in the lower left. Increasingly negative $V_\mathrm{TG}$ depletes the local carrier density restricting the current flow to a narrow constriction between the split gates before eventual pinch-off.}
    \label{fig:qt1}
\end{figure*}

Having demonstrated material- and device-level improvements from nano-ironing, we next highlight its potential for enabling quantum transport studies in vdW heterostructures. 
A key requirement for quantum transport is to create a homogeneous two-dimensional electron gas (2DEG).\cite{zhang2009, Martin2008a, Goh2020,Liu2019a} This homogeneity needs atomically clean interfaces free from residual induced disorder to preserve high carrier mobility and enable experimental control over the local electrostatic potential and carrier density distribution. Such independent electrical control is necessary for precise confinement and tuning of the 2DEG, important for quantum information processing applications.\cite{Hanson2007,Liu2019a,Montblanch2023,DeLeon2021a,Lau2021,Lau2019a} For example in qubits, where state preparation, readout, and manipulation are performed with local electrostatic control.

Quantum transport is therefore typically only observed in high quality heterostructures assembled before any lithographic process to minimize trapped resist residue between interfaces.\cite{Zhang2017i,Hamer2018,Wang2018,Pisoni2018,Eich2018,Davari2020,Boddison-Chouinard2021} However, this approach limits the choice of contact materials and introduces greater fabrication complexity.\cite{Wang2013,Pisoni2018,Wang2018,Davari2020,Boddison-Chouinard2021,Hamer2018} Complete reliance on such manually assembled contacts is intrinsically unscalable for technological applications. Another approach is to establish contacts post-heterostructure assembly with selective etching of the different vdW layers.\cite{Wang2013,Xu2016a} This requires a high degree of precision and can lead to unwanted damage to the channel surface. A direct, damage- and residue-free contact lithography process that can preserve vdW heterostructure interface cleanliness is desired to increase experimental flexibility, material choices, and yield. 

Here we demonstrate such a vdW heterostructure with atomically clean interfaces from nano-ironing (Figure \ref{fig:qt1}a). Our device consists of a hBN/WS$_2$/hBN stack with a graphitic back gate electrode to tune the overall 2DEG carrier density. Local electrostatic control is induced through two split top gates (labelled TG1 and TG2) with a separation of $\sim$\ 70 nm (Figure \ref{fig:qt1}b). Before the top hBN dielectric is transferred over the WS$_2$ channel, In/Au contacts are lithographically defined which leaves behind resist residue. Nano-ironing removes this resist residue before top hBN encapsulation resulting in a clean interface for cryogenic quantum transport measurements. 

We first assess the electrical transport characteristics of our vdW heterostructure at cryogenic temperatures. The transfer and output curves of the device measured at a temperature of 43 mK are shown in Figure \ref{fig:qt1}c where we observe typical $n$-type semiconducting behaviour. The high-quality of our In/Au contacts with negligibly low Schottky barrier heights are confirmed from the linear output characteristics.\cite{Lau2020,Zheng2019b} Correspondingly, we measure a low device resistance $R \sim$ 25 k$\Omega$, indicating that contacts are sufficiently transparent with contact resistances $R_\mathrm{C}<R/2$=12.5 k$\Omega$.

\begin{figure*}
    \centering
    \includegraphics[width=15cm]{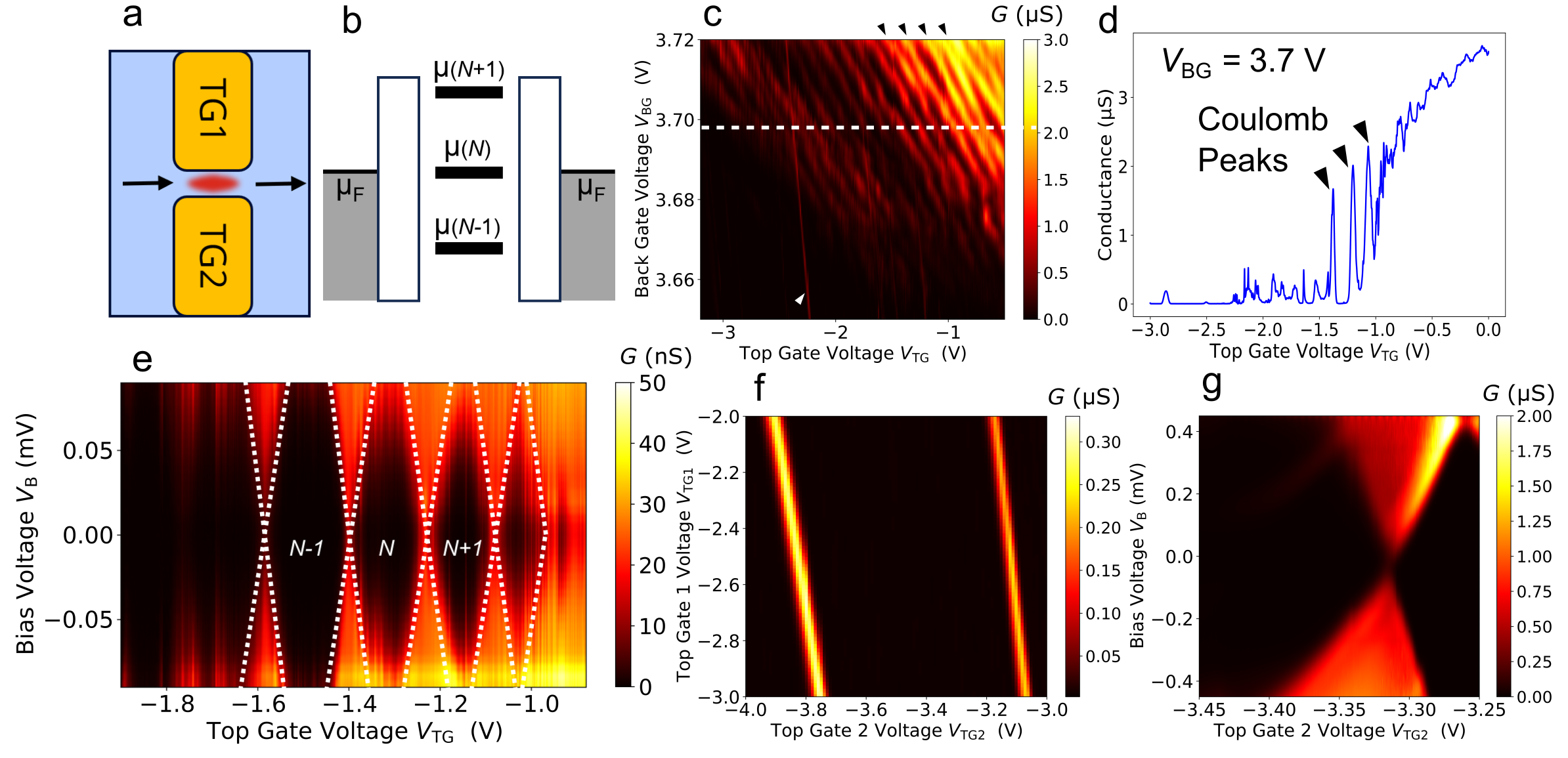}
    \caption{\textbf{Electrical formation and control of quantum dots}. (a) Schematic illustration of gate-induced quantum dot formation along the channel and (b) single-electron transport. The bias voltage defines the Fermi levels $\mu_\mathrm{F}$ of the left and right electrodes. A gate voltage controls the quantized electrochemical potential $\mu_{N}$ of the quantum dot transitions from an occupation state of $N$ electrons to $N$+1. Resonant single electron transport occurs when $\mu_{N}$ is aligned with $\mu_\mathrm{F}$ of the electrodes. Such resonances are observed in (c) which shows the device conductance $G$ as a function of the back gate voltage $V_\mathrm{BG}$ and a common top gate voltage $V_\mathrm{TG}$ applied to both split top gates. Several diagonal parallel resonances (black arrows) due to quantum dot formation along the constriction are observed. A vertical resonance with a different slope is also present (white arrow) suggestive of disorder induced accidental quantum dots localised under the top gate. (d) Device conductance $G$ as a function of top gate voltage $V_\mathrm{TG}$ at a fixed back gate voltage $V_\mathrm{BG}$=3.7 V (white horizontal dashed line in c) showing Coulomb peaks (black arrows) from resonant transport across the quantum dot. (e) Charge stability diagram showing diamond shaped regions of suppressed conductance indicative of Coulomb blockade at $V_\mathrm{BG}$ = 3.7 V. The diamond sizes increase with more negative $V_\mathrm{TG}$ due to a reduction in quantum dot size with stronger electrostatic confinement. (f) Device conductance $G$ as a function of $V_\mathrm{TG1}$ and $V_\mathrm{TG2}$ with $V_\mathrm{BG}$=3.6 V showing that the dot can be modulated by both split top gates. (g) Higher resolution stability diagram showing a Coulomb peak transition tuned by a single top gate 2 voltage $V_\mathrm{TG2}$ with $V_\mathrm{BG}$ = 3.6 V and $V_\mathrm{TG1}$= -1.5 V.} 
    \label{fig:qt2}
    
\end{figure*}

Next, we evaluate the electrical tunability of our device. We measure the device conductance $G$ with standard low frequency lock-in measurements as a function of the back gate voltage $V_\mathrm{BG}$ and a common top gate voltage $V_\mathrm{TG}$ applied to both split top gates TG1 and TG2 (Figure \ref{fig:qt1}d). The white symbols in Figure \ref{fig:qt1}d indicate the transport regimes described below and in the inset schematics. We find that $G$ can be effectively modulated by both $V_\mathrm{BG}$ and $V_\mathrm{TG}$. A more positive $V_\mathrm{BG}$ increases the overall carrier density and $G$. At small negative $V_\mathrm{TG}$, the electrostatic modulation is insufficient to deplete the local carrier densities underneath the gates resulting in current flow (star). With increasingly negative $V_\mathrm{TG}$, the local carrier densities are sufficiently depleted until current flow is largely restricted along the constriction between the split gates (pentagon), before complete pinch-off (circle). 

Having shown dual gate control, we next demonstrate the electrical formation of quantum dots. Negative voltages on the nanoscale split top gates can create electrostatic potential wells that confine islands of charged carriers, i.e., quantum dots (Figure \ref{fig:qt2}a). Such quantum confinement is only possible in 2DEGs with a low degree of disorder. Otherwise, the disorder can dominate transport through the creation of highly untunable accidental quantum dots.\cite{Liu2010,Lee2016,Zhang2017i,Brotons-Gisbert2019d} Quantum dot formation results in a transition from macroscopic current flow to the single-electron transport regime, where electron-electron interactions lead to Coulomb blockade.\cite{Hanson2007} In this regime, transport across the device is Coulomb blocked unless an electrochemical potential level of the quantum dot is in resonance with the Fermi levels of the electrodes (Figure \ref{fig:qt2}b), which then manifest as Coulomb conductance peaks. 

Tunability of an electrically defined quantum dot by both $V_\mathrm{BG}$ and $V_\mathrm{TG}$ indicates their formation in the narrow constriction between the split top gates. Indeed, we observe such resonances in the form of several diagonal lines (indicated by the black arrows) in Figure \ref{fig:qt2}c. Other resonances can also be observed (white arrow). As mentioned, these are likely due to accidental disorder-induced quantum dots located directly under a top gate rather than in the narrow constriction, and so appears as a nearly vertical line largely tunable only by $V_\mathrm{TG}$. The transition from macroscopic current flow to Coulomb blockade can be visualised from Figure \ref{fig:qt2}d which shows 1D $G$ sweeps. Strong Coulomb conductance peaks are observed as expected (black arrows). 

The clearest indication of Coulomb blockade is obtained from transport spectroscopy of stability diagrams where $G$ is measured while changing the bias voltage $V_\mathrm{B}$ and $V_\mathrm{TG}$ (Figure \ref{fig:qt2}e). With increasingly negative $V_\mathrm{TG}$, we observe the appearance of consecutively larger diamond-shaped regions of suppressed conductance. This is due to Coulomb blockade where no electrochemical potential levels of the quantum dot lie within the applied bias window and the total charge $N$ on the quantum dot is stable. The heights of these Coulomb diamonds along the $V_\mathrm{B}$ axis give the charging energy $E_\mathrm{C}$ of the quantum dot. $E_\mathrm{C}$ is the energy required to add/remove an electron from the island arising from electron-electron Coulombic repulsion and is inversely proportional to the quantum dot radius $r$.\cite{Hanson2007,Lau2021} The increase in Coulomb diamond heights with increasingly negative $V_\mathrm{TG}$ is thus consistent with a decrease in the quantum dot size from stronger electrostatic confinement. 

We further confirm gate tunability of the quantum dots with independent top gate sweeps of gates TG1 and TG2. We again observe diagonal Coulomb peak resonances in Figure \ref{fig:qt2}f which shows $G$ as a function of $V_\mathrm{TG1}$ and $V_\mathrm{TG2}$. The resonance slope values are $\sim$5, indicating that the dot is more strongly coupled to TG2 than TG1. Using only TG2, we can also observe a Coulomb peak transition in the bias stability diagram (Figure \ref{fig:qt2}g). However, we note that the single gate tunability range is limited as it also results in a more asymmetric confinement potential. Future devices implementing additional independent gates for improved control over tunnel couplings should open up wider possibilities for 2D material-based quantum devices. 

Beyond transport in nano- and quantum electronics, nano-ironing can be useful in many other applications. For example, reducing subthreshold swings is important for fast, low-power transistors but gate capacitances can be limited by resist residue.\cite{Zhang2023,Lau2023} Nano-ironing can lead to fast transistors with subthreshold swings approaching the thermionic limit of 60 mV/dec.\cite{Das2021} In opto-electronics, resist residual removal improves photon absorption (emission) for photo-detectors (photo-emitters).\cite{Tan2016,Tan2017c} Excitonic resonances in 2D TMDs are sensitive to the dielectric environment which can be dominated by resist residue.\cite{Raja2017} Nano-ironing can improve exciton lifetimes and yields including inter-layer excitonic species in vdW heterostructures. In bio-sensing applications, device sensitivity and specificity can benefit from increased useful surface areas for detection or functionalization.\cite{Sarkar2014,Lau2023} Valleytronic and spintronic devices can expect improved valley/spin lifetimes and diffusion lengths from reduced roughness and impurity scattering with cleaner vdW surfaces.\cite{Goh2020,Avsar2020,Schaibley2016} Proximity induced effects such as anomalous Hall effect\cite{Wang2015} and Zeeman Spin Hall effect\cite{Wei2016} are also strongly dependent on interface cleanliness and can benefit from nano-ironing. 

In summary, we show that a heated nano-probe tip efficiently removes residual resist (nano-ironing). This reduces surface roughness and recovers an atomically flat vdW material surface. Nano-ironed monolayer WS$_2$ transistors display improved electrical performance with reduced unwanted charge doping, higher carrier mobilities, and higher drain currents. Nano-ironing is useful for building vdW heterostructures with ultra-clean interfaces to create homogeneous 2DEGs for cryogenic quantum transport studies. We demonstrate electrical confinement and control of quantum dots in nano-ironed vdW heterostructures, enabling the transition from macroscopic current flow to single-electron transport regime. Such vdW heterostructure based quantum dots are useful platforms to unveil fundamental transport physics and important building blocks for quantum information processing applications. Nano-ironing thus paves the way towards new fabrication approaches and device architectures for vdW heterostructures.

\section{Materials and Methods}

\subsection{Thermal scanning probe for lithography and nano-ironing}

The t-SPL system used for both contact patterning and nano-ironing was a Nanofrazor Scholar from Heidelberg Instruments kept in an inert glovebox. For t-SPL patterning procedure we spincoat a bi-layer resist stack consisting of 30/110 nm of polyphthalaldehyde (PPA) / poly (methyl methacrylate-co-methacrylic acid) (PMMA-MA) at 4000 rpm for 40 seconds. Samples are baked at 140 \textsuperscript{$\circ$}C for 90 seconds after spinning the PMMA-MA layer, and at 110 \textsuperscript{$\circ$}C for 120 seconds after spinning the PPA layer. After patterning, we etched the exposed PMMA-MA using 20 \textsuperscript{$\circ$}C ethanol for 15 seconds. For t-SPL nano-ironing, a heater temperature of 1100 \textsuperscript{$\circ$}C, a write force of 7.5 V, and a force pulse of 450 microseconds per pixel is used.

\subsection{Device fabrication}

Device fabrication were conducted in an inert glovebox unless otherwise stated. CVD grown monolayer WS$_\mathrm{2}$ was transferred onto the Si/SiO$_\mathrm{2}$ device substrate via a PPC assisted transfer method.\cite{Mondal2023} The transfer method involves preparation steps which must be done outside of the glovebox. However, the final transfer of the CVD WS$_2$ onto Si/SiO$_\mathrm{2}$ is performed inside the glovebox. 5/35 nm In/Au contacts were defined via: (i) thermal scanning probe lithography (t-SPL) (ii) thermal evaporation (iii) lift-off. 2/50 nm Cr/Au lines and pads were then defined via: (i) Electron beam lithography (EBL) (ii) thermal evaporation (iii) lift-off. Electron beam lithography was carried out outside of the glove-box. Samples were passivated with resist while being transported between systems. 

For the heterostructure device, we exfoliated graphite, hBN, and WS$_2$ flakes and stacked them with standard polymer-assisted dry transfer methods on an automated transfer stage setup.\cite{Mukherjee2023} After patterning the source, drain, and back-gate contacts with t-SPL, (channel length of 1.5 $\mathrm{\mu}$m with a width of $\sim$4 $\mathrm{\mu}$m), we deposited 5/35 nm of In/Au alloy via thermal evaporator. After lift-off, we conducted nano-ironing on the channel to clean the interface, then used the same stacking system to encapsulate the channel with a top layer of hBN. We patterned the split gates (separated by $\sim$70 nm with a width of $\sim$800 nm) through EBL followed by Cr/Au (3/22 nm) metal deposition. Subsequent lines and pads were also defined by EBL, followed by 5/55 nm of Cr/Au deposition. 

Further details of nanofabrication recipes and processes can be found in the Supporting Information.

\subsection{FET transport measurements}
FET transport measurements were conducted in a vacuum cryoprobe station cooled down to 80 K at a base pressure of $\sim$10\textsuperscript{-6} Torr. All electrical measurements were carried out with a Keithley 2450 source meter. 

\subsection{Quantum transport measurements}
Quantum transport measurements were conducted in a BlueFors XLD dilution refrigerator at a base temperature of 43 mK using a Nanonis Tramea measurement system. 

\subsection{Optical measurements}
Room temperature PL and Raman spectroscopy were done through Invia Raman Renishaw system. The point spectra were obtained using a 532 nm laser with $\times$100 objective (NA = 0.85) and 2400 $l$/mm grating. The excitation power was kept below 50 $\mu$W to avoid any unintentional degradation due to laser-induced heating. The mapping was performed by scanning the sample in 7 $\times$ 7 $\mu$m area through the same laser excitation. 

\begin{acknowledgements}
This research was supported by the Agency for Science, Technology, and Research (A*STAR) under its MTC YIRG grant No. M21K3c0124. We acknowledge the funding support from the Agency for Science, Technology and Research (\#21709, C230917006, C230917007). K.E.J.G. acknowledges support from a Singapore National Research Foundation Grant (CRP21-2018-0001). D.H. acknowledges funding support from A*STAR Project C222812022 and MTC YIRG M22K3c0105.
\end{acknowledgements}

\bibliography{Ref}

\begin{thebibliography}{58}%
\makeatletter
\providecommand \@ifxundefined [1]{%
 \@ifx{#1\undefined}
}%
\providecommand \@ifnum [1]{%
 \ifnum #1\expandafter \@firstoftwo
 \else \expandafter \@secondoftwo
 \fi
}%
\providecommand \@ifx [1]{%
 \ifx #1\expandafter \@firstoftwo
 \else \expandafter \@secondoftwo
 \fi
}%
\providecommand \natexlab [1]{#1}%
\providecommand \enquote  [1]{``#1''}%
\providecommand \bibnamefont  [1]{#1}%
\providecommand \bibfnamefont [1]{#1}%
\providecommand \citenamefont [1]{#1}%
\providecommand \href@noop [0]{\@secondoftwo}%
\providecommand \href [0]{\begingroup \@sanitize@url \@href}%
\providecommand \@href[1]{\@@startlink{#1}\@@href}%
\providecommand \@@href[1]{\endgroup#1\@@endlink}%
\providecommand \@sanitize@url [0]{\catcode `\\12\catcode `\$12\catcode `\&12\catcode `\#12\catcode `\^12\catcode `\_12\catcode `\%12\relax}%
\providecommand \@@startlink[1]{}%
\providecommand \@@endlink[0]{}%
\providecommand \url  [0]{\begingroup\@sanitize@url \@url }%
\providecommand \@url [1]{\endgroup\@href {#1}{\urlprefix }}%
\providecommand \urlprefix  [0]{URL }%
\providecommand \Eprint [0]{\href }%
\providecommand \doibase [0]{https://doi.org/}%
\providecommand \selectlanguage [0]{\@gobble}%
\providecommand \bibinfo  [0]{\@secondoftwo}%
\providecommand \bibfield  [0]{\@secondoftwo}%
\providecommand \translation [1]{[#1]}%
\providecommand \BibitemOpen [0]{}%
\providecommand \bibitemStop [0]{}%
\providecommand \bibitemNoStop [0]{.\EOS\space}%
\providecommand \EOS [0]{\spacefactor3000\relax}%
\providecommand \BibitemShut  [1]{\csname bibitem#1\endcsname}%
\let\auto@bib@innerbib\@empty
%</preamble>
\bibitem [{\citenamefont {Novoselov}\ \emph {et~al.}(2004)\citenamefont {Novoselov}, \citenamefont {Geim}, \citenamefont {Morozov}, \citenamefont {Jiang}, \citenamefont {Zhang}, \citenamefont {Dubonos}, \citenamefont {Grigorieva},\ and\ \citenamefont {Firsov}}]{Novoselov2004}%
  \BibitemOpen
  \bibfield  {author} {\bibinfo {author} {\bibfnamefont {K.~S.}\ \bibnamefont {Novoselov}}, \bibinfo {author} {\bibfnamefont {A.~K.}\ \bibnamefont {Geim}}, \bibinfo {author} {\bibfnamefont {S.~V.}\ \bibnamefont {Morozov}}, \bibinfo {author} {\bibfnamefont {D.}~\bibnamefont {Jiang}}, \bibinfo {author} {\bibfnamefont {Y.}~\bibnamefont {Zhang}}, \bibinfo {author} {\bibfnamefont {S.~V.}\ \bibnamefont {Dubonos}}, \bibinfo {author} {\bibfnamefont {I.~V.}\ \bibnamefont {Grigorieva}},\ and\ \bibinfo {author} {\bibfnamefont {A.~A.}\ \bibnamefont {Firsov}},\ }\bibfield  {title} {\bibinfo {title} {Electric field effect in atomically thin carbon films},\ }\href@noop {} {\bibfield  {journal} {\bibinfo  {journal} {Science}\ }\textbf {\bibinfo {volume} {306}},\ \bibinfo {pages} {666} (\bibinfo {year} {2004})}\BibitemShut {NoStop}%
\bibitem [{\citenamefont {Haigh}\ \emph {et~al.}(2012)\citenamefont {Haigh}, \citenamefont {Gholinia}, \citenamefont {Jalil}, \citenamefont {Romani}, \citenamefont {Britnell}, \citenamefont {Elias}, \citenamefont {Novoselov}, \citenamefont {Ponomarenko}, \citenamefont {Geim},\ and\ \citenamefont {Gorbachev}}]{Haigh2012a}%
  \BibitemOpen
  \bibfield  {author} {\bibinfo {author} {\bibfnamefont {S.~J.}\ \bibnamefont {Haigh}}, \bibinfo {author} {\bibfnamefont {A.}~\bibnamefont {Gholinia}}, \bibinfo {author} {\bibfnamefont {R.}~\bibnamefont {Jalil}}, \bibinfo {author} {\bibfnamefont {S.}~\bibnamefont {Romani}}, \bibinfo {author} {\bibfnamefont {L.}~\bibnamefont {Britnell}}, \bibinfo {author} {\bibfnamefont {D.~C.}\ \bibnamefont {Elias}}, \bibinfo {author} {\bibfnamefont {K.~S.}\ \bibnamefont {Novoselov}}, \bibinfo {author} {\bibfnamefont {L.~A.}\ \bibnamefont {Ponomarenko}}, \bibinfo {author} {\bibfnamefont {A.~K.}\ \bibnamefont {Geim}},\ and\ \bibinfo {author} {\bibfnamefont {R.}~\bibnamefont {Gorbachev}},\ }\bibfield  {title} {\bibinfo {title} {{Cross-sectional imaging of individual layers and buried interfaces of graphene-based heterostructures and superlattices}},\ }\href@noop {} {\bibfield  {journal} {\bibinfo  {journal} {Nature Materials}\ }\textbf {\bibinfo {volume} {11}},\ \bibinfo {pages} {764} (\bibinfo {year} {2012})}\BibitemShut
  {NoStop}%
\bibitem [{\citenamefont {Geim}\ and\ \citenamefont {Grigorieva}(2013)}]{Geim2013a}%
  \BibitemOpen
  \bibfield  {author} {\bibinfo {author} {\bibfnamefont {A.~K.}\ \bibnamefont {Geim}}\ and\ \bibinfo {author} {\bibfnamefont {I.~V.}\ \bibnamefont {Grigorieva}},\ }\bibfield  {title} {\bibinfo {title} {{Van der Waals heterostructures}},\ }\href@noop {} {\bibfield  {journal} {\bibinfo  {journal} {Nature}\ }\textbf {\bibinfo {volume} {499}},\ \bibinfo {pages} {419} (\bibinfo {year} {2013})}\BibitemShut {NoStop}%
\bibitem [{\citenamefont {Pizzocchero}\ \emph {et~al.}(2016)\citenamefont {Pizzocchero}, \citenamefont {Gammelgaard}, \citenamefont {Jessen}, \citenamefont {Caridad}, \citenamefont {Wang}, \citenamefont {Hone}, \citenamefont {B{\o}ggild},\ and\ \citenamefont {Booth}}]{Pizzocchero2016}%
  \BibitemOpen
  \bibfield  {author} {\bibinfo {author} {\bibfnamefont {F.}~\bibnamefont {Pizzocchero}}, \bibinfo {author} {\bibfnamefont {L.}~\bibnamefont {Gammelgaard}}, \bibinfo {author} {\bibfnamefont {B.~S.}\ \bibnamefont {Jessen}}, \bibinfo {author} {\bibfnamefont {J.~M.}\ \bibnamefont {Caridad}}, \bibinfo {author} {\bibfnamefont {L.}~\bibnamefont {Wang}}, \bibinfo {author} {\bibfnamefont {J.}~\bibnamefont {Hone}}, \bibinfo {author} {\bibfnamefont {P.}~\bibnamefont {B{\o}ggild}},\ and\ \bibinfo {author} {\bibfnamefont {T.~J.}\ \bibnamefont {Booth}},\ }\bibfield  {title} {\bibinfo {title} {{The hot pick-up technique for batch assembly of van der Waals heterostructures}},\ }\href@noop {} {\bibfield  {journal} {\bibinfo  {journal} {Nature Communications}\ }\textbf {\bibinfo {volume} {7}},\ \bibinfo {pages} {11894} (\bibinfo {year} {2016})}\BibitemShut {NoStop}%
\bibitem [{\citenamefont {Cao}\ \emph {et~al.}(2018)\citenamefont {Cao}, \citenamefont {Fatemi}, \citenamefont {Fang}, \citenamefont {Watanabe}, \citenamefont {Taniguchi}, \citenamefont {Kaxiras},\ and\ \citenamefont {Jarillo-Herrero}}]{Cao2018a}%
  \BibitemOpen
  \bibfield  {author} {\bibinfo {author} {\bibfnamefont {Y.}~\bibnamefont {Cao}}, \bibinfo {author} {\bibfnamefont {V.}~\bibnamefont {Fatemi}}, \bibinfo {author} {\bibfnamefont {S.}~\bibnamefont {Fang}}, \bibinfo {author} {\bibfnamefont {K.}~\bibnamefont {Watanabe}}, \bibinfo {author} {\bibfnamefont {T.}~\bibnamefont {Taniguchi}}, \bibinfo {author} {\bibfnamefont {E.}~\bibnamefont {Kaxiras}},\ and\ \bibinfo {author} {\bibfnamefont {P.}~\bibnamefont {Jarillo-Herrero}},\ }\bibfield  {title} {\bibinfo {title} {{Unconventional superconductivity in magic-angle graphene superlattices}},\ }\href@noop {} {\bibfield  {journal} {\bibinfo  {journal} {Nature}\ }\textbf {\bibinfo {volume} {556}},\ \bibinfo {pages} {43} (\bibinfo {year} {2018})}\BibitemShut {NoStop}%
\bibitem [{\citenamefont {Sharpe}\ \emph {et~al.}(2019)\citenamefont {Sharpe}, \citenamefont {Fox}, \citenamefont {Barnard}, \citenamefont {Finney}, \citenamefont {Watanabe}, \citenamefont {Taniguchi}, \citenamefont {Kastner},\ and\ \citenamefont {Goldhaber-Gordon}}]{Sharpe2019}%
  \BibitemOpen
  \bibfield  {author} {\bibinfo {author} {\bibfnamefont {A.~L.}\ \bibnamefont {Sharpe}}, \bibinfo {author} {\bibfnamefont {E.~J.}\ \bibnamefont {Fox}}, \bibinfo {author} {\bibfnamefont {A.~W.}\ \bibnamefont {Barnard}}, \bibinfo {author} {\bibfnamefont {J.}~\bibnamefont {Finney}}, \bibinfo {author} {\bibfnamefont {K.}~\bibnamefont {Watanabe}}, \bibinfo {author} {\bibfnamefont {T.}~\bibnamefont {Taniguchi}}, \bibinfo {author} {\bibfnamefont {M.~A.}\ \bibnamefont {Kastner}},\ and\ \bibinfo {author} {\bibfnamefont {D.}~\bibnamefont {Goldhaber-Gordon}},\ }\bibfield  {title} {\bibinfo {title} {Emergent ferromagnetism near three-quarters filling in twisted bilayer graphene},\ }\href@noop {} {\bibfield  {journal} {\bibinfo  {journal} {Science}\ }\textbf {\bibinfo {volume} {365}},\ \bibinfo {pages} {605} (\bibinfo {year} {2019})}\BibitemShut {NoStop}%
\bibitem [{\citenamefont {Zheng}\ \emph {et~al.}(2020)\citenamefont {Zheng}, \citenamefont {Ma}, \citenamefont {Bi}, \citenamefont {de~La~Barrera}, \citenamefont {Liu}, \citenamefont {Mao}, \citenamefont {Zhang}, \citenamefont {Kiper}, \citenamefont {Watanabe}, \citenamefont {Taniguchi} \emph {et~al.}}]{Zheng2020}%
  \BibitemOpen
  \bibfield  {author} {\bibinfo {author} {\bibfnamefont {Z.}~\bibnamefont {Zheng}}, \bibinfo {author} {\bibfnamefont {Q.}~\bibnamefont {Ma}}, \bibinfo {author} {\bibfnamefont {Z.}~\bibnamefont {Bi}}, \bibinfo {author} {\bibfnamefont {S.}~\bibnamefont {de~La~Barrera}}, \bibinfo {author} {\bibfnamefont {M.-H.}\ \bibnamefont {Liu}}, \bibinfo {author} {\bibfnamefont {N.}~\bibnamefont {Mao}}, \bibinfo {author} {\bibfnamefont {Y.}~\bibnamefont {Zhang}}, \bibinfo {author} {\bibfnamefont {N.}~\bibnamefont {Kiper}}, \bibinfo {author} {\bibfnamefont {K.}~\bibnamefont {Watanabe}}, \bibinfo {author} {\bibfnamefont {T.}~\bibnamefont {Taniguchi}}, \emph {et~al.},\ }\bibfield  {title} {\bibinfo {title} {Unconventional ferroelectricity in moir{\'e} heterostructures},\ }\href@noop {} {\bibfield  {journal} {\bibinfo  {journal} {Nature}\ }\textbf {\bibinfo {volume} {588}},\ \bibinfo {pages} {71} (\bibinfo {year} {2020})}\BibitemShut {NoStop}%
\bibitem [{\citenamefont {Serlin}\ \emph {et~al.}(2020)\citenamefont {Serlin}, \citenamefont {Tschirhart}, \citenamefont {Polshyn}, \citenamefont {Zhang}, \citenamefont {Zhu}, \citenamefont {Watanabe}, \citenamefont {Taniguchi}, \citenamefont {Balents},\ and\ \citenamefont {Young}}]{Serlin2020}%
  \BibitemOpen
  \bibfield  {author} {\bibinfo {author} {\bibfnamefont {M.}~\bibnamefont {Serlin}}, \bibinfo {author} {\bibfnamefont {C.~L.}\ \bibnamefont {Tschirhart}}, \bibinfo {author} {\bibfnamefont {H.}~\bibnamefont {Polshyn}}, \bibinfo {author} {\bibfnamefont {Y.}~\bibnamefont {Zhang}}, \bibinfo {author} {\bibfnamefont {J.}~\bibnamefont {Zhu}}, \bibinfo {author} {\bibfnamefont {K.}~\bibnamefont {Watanabe}}, \bibinfo {author} {\bibfnamefont {T.}~\bibnamefont {Taniguchi}}, \bibinfo {author} {\bibfnamefont {L.}~\bibnamefont {Balents}},\ and\ \bibinfo {author} {\bibfnamefont {A.~F.}\ \bibnamefont {Young}},\ }\bibfield  {title} {\bibinfo {title} {Intrinsic quantized anomalous hall effect in a moiré heterostructure},\ }\href@noop {} {\bibfield  {journal} {\bibinfo  {journal} {Science}\ }\textbf {\bibinfo {volume} {367}},\ \bibinfo {pages} {900} (\bibinfo {year} {2020})}\BibitemShut {NoStop}%
\bibitem [{\citenamefont {Dan}\ \emph {et~al.}(2009)\citenamefont {Dan}, \citenamefont {Lu}, \citenamefont {Kybert}, \citenamefont {Luo},\ and\ \citenamefont {Johnson}}]{Dan2009}%
  \BibitemOpen
  \bibfield  {author} {\bibinfo {author} {\bibfnamefont {Y.}~\bibnamefont {Dan}}, \bibinfo {author} {\bibfnamefont {Y.}~\bibnamefont {Lu}}, \bibinfo {author} {\bibfnamefont {N.~J.}\ \bibnamefont {Kybert}}, \bibinfo {author} {\bibfnamefont {Z.}~\bibnamefont {Luo}},\ and\ \bibinfo {author} {\bibfnamefont {A.~T.~C.}\ \bibnamefont {Johnson}},\ }\bibfield  {title} {\bibinfo {title} {Intrinsic response of graphene vapor sensors},\ }\href@noop {} {\bibfield  {journal} {\bibinfo  {journal} {Nano Letters}\ }\textbf {\bibinfo {volume} {9}},\ \bibinfo {pages} {1472} (\bibinfo {year} {2009})}\BibitemShut {NoStop}%
\bibitem [{\citenamefont {Cheng}\ \emph {et~al.}(2011)\citenamefont {Cheng}, \citenamefont {Zhou}, \citenamefont {Wang}, \citenamefont {Li}, \citenamefont {Wang},\ and\ \citenamefont {Fang}}]{Cheng2011}%
  \BibitemOpen
  \bibfield  {author} {\bibinfo {author} {\bibfnamefont {Z.}~\bibnamefont {Cheng}}, \bibinfo {author} {\bibfnamefont {Q.}~\bibnamefont {Zhou}}, \bibinfo {author} {\bibfnamefont {C.}~\bibnamefont {Wang}}, \bibinfo {author} {\bibfnamefont {Q.}~\bibnamefont {Li}}, \bibinfo {author} {\bibfnamefont {C.}~\bibnamefont {Wang}},\ and\ \bibinfo {author} {\bibfnamefont {Y.}~\bibnamefont {Fang}},\ }\bibfield  {title} {\bibinfo {title} {Toward intrinsic graphene surfaces: A systematic study on thermal annealing and wet-chemical treatment of sio2-supported graphene devices},\ }\href@noop {} {\bibfield  {journal} {\bibinfo  {journal} {Nano Letters}\ }\textbf {\bibinfo {volume} {11}},\ \bibinfo {pages} {767} (\bibinfo {year} {2011})}\BibitemShut {NoStop}%
\bibitem [{\citenamefont {Bolotin}\ \emph {et~al.}(2008)\citenamefont {Bolotin}, \citenamefont {Sikes}, \citenamefont {Jiang}, \citenamefont {Klima}, \citenamefont {Fudenberg}, \citenamefont {Hone}, \citenamefont {Kim},\ and\ \citenamefont {Stormer}}]{Bolotin2008}%
  \BibitemOpen
  \bibfield  {author} {\bibinfo {author} {\bibfnamefont {K.}~\bibnamefont {Bolotin}}, \bibinfo {author} {\bibfnamefont {K.}~\bibnamefont {Sikes}}, \bibinfo {author} {\bibfnamefont {Z.}~\bibnamefont {Jiang}}, \bibinfo {author} {\bibfnamefont {M.}~\bibnamefont {Klima}}, \bibinfo {author} {\bibfnamefont {G.}~\bibnamefont {Fudenberg}}, \bibinfo {author} {\bibfnamefont {J.}~\bibnamefont {Hone}}, \bibinfo {author} {\bibfnamefont {P.}~\bibnamefont {Kim}},\ and\ \bibinfo {author} {\bibfnamefont {H.}~\bibnamefont {Stormer}},\ }\bibfield  {title} {\bibinfo {title} {Ultrahigh electron mobility in suspended graphene},\ }\href@noop {} {\bibfield  {journal} {\bibinfo  {journal} {Solid State Communications}\ }\textbf {\bibinfo {volume} {146}},\ \bibinfo {pages} {351} (\bibinfo {year} {2008})}\BibitemShut {NoStop}%
\bibitem [{\citenamefont {Lindvall}\ \emph {et~al.}(2012)\citenamefont {Lindvall}, \citenamefont {Kalabukhov},\ and\ \citenamefont {Yurgens}}]{Lindvall2012a}%
  \BibitemOpen
  \bibfield  {author} {\bibinfo {author} {\bibfnamefont {N.}~\bibnamefont {Lindvall}}, \bibinfo {author} {\bibfnamefont {A.}~\bibnamefont {Kalabukhov}},\ and\ \bibinfo {author} {\bibfnamefont {A.}~\bibnamefont {Yurgens}},\ }\bibfield  {title} {\bibinfo {title} {{Cleaning graphene using atomic force microscope}},\ }\href@noop {} {\bibfield  {journal} {\bibinfo  {journal} {Journal of Applied Physics}\ }\textbf {\bibinfo {volume} {111}},\ \bibinfo {pages} {64904} (\bibinfo {year} {2012})}\BibitemShut {NoStop}%
\bibitem [{\citenamefont {Rosenberger}\ \emph {et~al.}(2018)\citenamefont {Rosenberger}, \citenamefont {Chuang}, \citenamefont {McCreary}, \citenamefont {Hanbicki}, \citenamefont {Sivaram},\ and\ \citenamefont {Jonker}}]{Rosenberger2018}%
  \BibitemOpen
  \bibfield  {author} {\bibinfo {author} {\bibfnamefont {M.~R.}\ \bibnamefont {Rosenberger}}, \bibinfo {author} {\bibfnamefont {H.~J.}\ \bibnamefont {Chuang}}, \bibinfo {author} {\bibfnamefont {K.~M.}\ \bibnamefont {McCreary}}, \bibinfo {author} {\bibfnamefont {A.~T.}\ \bibnamefont {Hanbicki}}, \bibinfo {author} {\bibfnamefont {S.~V.}\ \bibnamefont {Sivaram}},\ and\ \bibinfo {author} {\bibfnamefont {B.~T.}\ \bibnamefont {Jonker}},\ }\bibfield  {title} {\bibinfo {title} {{Nano-"Squeegee" for the Creation of Clean 2D Material Interfaces}},\ }\href@noop {} {\bibfield  {journal} {\bibinfo  {journal} {ACS Applied Materials and Interfaces}\ }\textbf {\bibinfo {volume} {10}},\ \bibinfo {pages} {10379} (\bibinfo {year} {2018})}\BibitemShut {NoStop}%
\bibitem [{\citenamefont {Kim}\ \emph {et~al.}(2019)\citenamefont {Kim}, \citenamefont {Herlinger}, \citenamefont {Taniguchi}, \citenamefont {Watanabe},\ and\ \citenamefont {Smet}}]{Kim2019c}%
  \BibitemOpen
  \bibfield  {author} {\bibinfo {author} {\bibfnamefont {Y.}~\bibnamefont {Kim}}, \bibinfo {author} {\bibfnamefont {P.}~\bibnamefont {Herlinger}}, \bibinfo {author} {\bibfnamefont {T.}~\bibnamefont {Taniguchi}}, \bibinfo {author} {\bibfnamefont {K.}~\bibnamefont {Watanabe}},\ and\ \bibinfo {author} {\bibfnamefont {J.~H.}\ \bibnamefont {Smet}},\ }\bibfield  {title} {\bibinfo {title} {{Reliable Postprocessing Improvement of van der Waals Heterostructures}},\ }\href@noop {} {\bibfield  {journal} {\bibinfo  {journal} {ACS Nano}\ }\textbf {\bibinfo {volume} {13}},\ \bibinfo {pages} {14182} (\bibinfo {year} {2019})}\BibitemShut {NoStop}%
\bibitem [{\citenamefont {Palai}\ \emph {et~al.}(2023)\citenamefont {Palai}, \citenamefont {Dyksik}, \citenamefont {Sokolowski}, \citenamefont {Ciorga}, \citenamefont {{S{\'{a}}nchez Viso}}, \citenamefont {Xie}, \citenamefont {Schubert}, \citenamefont {Taniguchi}, \citenamefont {Watanabe}, \citenamefont {Maude}, \citenamefont {Surrente}, \citenamefont {Baranowski}, \citenamefont {Castellanos-Gomez}, \citenamefont {Munuera},\ and\ \citenamefont {Plochocka}}]{Palai2023}%
  \BibitemOpen
  \bibfield  {author} {\bibinfo {author} {\bibfnamefont {S.~K.}\ \bibnamefont {Palai}}, \bibinfo {author} {\bibfnamefont {M.}~\bibnamefont {Dyksik}}, \bibinfo {author} {\bibfnamefont {N.}~\bibnamefont {Sokolowski}}, \bibinfo {author} {\bibfnamefont {M.}~\bibnamefont {Ciorga}}, \bibinfo {author} {\bibfnamefont {E.}~\bibnamefont {{S{\'{a}}nchez Viso}}}, \bibinfo {author} {\bibfnamefont {Y.}~\bibnamefont {Xie}}, \bibinfo {author} {\bibfnamefont {A.}~\bibnamefont {Schubert}}, \bibinfo {author} {\bibfnamefont {T.}~\bibnamefont {Taniguchi}}, \bibinfo {author} {\bibfnamefont {K.}~\bibnamefont {Watanabe}}, \bibinfo {author} {\bibfnamefont {D.~K.}\ \bibnamefont {Maude}}, \bibinfo {author} {\bibfnamefont {A.}~\bibnamefont {Surrente}}, \bibinfo {author} {\bibfnamefont {M.}~\bibnamefont {Baranowski}}, \bibinfo {author} {\bibfnamefont {A.}~\bibnamefont {Castellanos-Gomez}}, \bibinfo {author} {\bibfnamefont {C.}~\bibnamefont {Munuera}},\ and\ \bibinfo {author} {\bibfnamefont {P.}~\bibnamefont {Plochocka}},\ }\bibfield
  {title} {\bibinfo {title} {{Approaching the Intrinsic Properties of Moir{\'{e}} Structures Using Atomic Force Microscopy Ironing.}},\ }\href@noop {} {\bibfield  {journal} {\bibinfo  {journal} {Nano Letters}\ }\textbf {\bibinfo {volume} {23}},\ \bibinfo {pages} {4749} (\bibinfo {year} {2023})}\BibitemShut {NoStop}%
\bibitem [{\citenamefont {Liu}\ \emph {et~al.}(2021)\citenamefont {Liu}, \citenamefont {Duan}, \citenamefont {Shin}, \citenamefont {Park}, \citenamefont {Huang},\ and\ \citenamefont {Duan}}]{Liu2021}%
  \BibitemOpen
  \bibfield  {author} {\bibinfo {author} {\bibfnamefont {Y.}~\bibnamefont {Liu}}, \bibinfo {author} {\bibfnamefont {X.}~\bibnamefont {Duan}}, \bibinfo {author} {\bibfnamefont {H.-J.}\ \bibnamefont {Shin}}, \bibinfo {author} {\bibfnamefont {S.}~\bibnamefont {Park}}, \bibinfo {author} {\bibfnamefont {Y.}~\bibnamefont {Huang}},\ and\ \bibinfo {author} {\bibfnamefont {X.}~\bibnamefont {Duan}},\ }\bibfield  {title} {\bibinfo {title} {{Promises and prospects of two-dimensional transistors}},\ }\href@noop {} {\bibfield  {journal} {\bibinfo  {journal} {Nature}\ }\textbf {\bibinfo {volume} {591}},\ \bibinfo {pages} {43} (\bibinfo {year} {2021})}\BibitemShut {NoStop}%
\bibitem [{\citenamefont {Lau}\ \emph {et~al.}(2014)\citenamefont {Lau}, \citenamefont {Mol}, \citenamefont {Warner},\ and\ \citenamefont {Briggs}}]{Lau2014}%
  \BibitemOpen
  \bibfield  {author} {\bibinfo {author} {\bibfnamefont {C.~S.}\ \bibnamefont {Lau}}, \bibinfo {author} {\bibfnamefont {J.~A.}\ \bibnamefont {Mol}}, \bibinfo {author} {\bibfnamefont {J.~H.}\ \bibnamefont {Warner}},\ and\ \bibinfo {author} {\bibfnamefont {G.~A.~D.}\ \bibnamefont {Briggs}},\ }\bibfield  {title} {\bibinfo {title} {{Nanoscale control of graphene electrodes}},\ }\href@noop {} {\bibfield  {journal} {\bibinfo  {journal} {Phys. Chem. Chem. Phys.}\ }\textbf {\bibinfo {volume} {16}},\ \bibinfo {pages} {20398} (\bibinfo {year} {2014})}\BibitemShut {NoStop}%
\bibitem [{\citenamefont {Zheng}\ \emph {et~al.}(2019{\natexlab{a}})\citenamefont {Zheng}, \citenamefont {Cal{\`{o}}}, \citenamefont {Albisetti}, \citenamefont {Liu}, \citenamefont {Alharbi}, \citenamefont {Arefe}, \citenamefont {Liu}, \citenamefont {Spieser}, \citenamefont {Yoo}, \citenamefont {Taniguchi}, \citenamefont {Watanabe}, \citenamefont {Aruta}, \citenamefont {Ciarrocchi}, \citenamefont {Kis}, \citenamefont {Lee}, \citenamefont {Lipson}, \citenamefont {Hone}, \citenamefont {Shahrjerdi},\ and\ \citenamefont {Riedo}}]{Zheng2019a}%
  \BibitemOpen
  \bibfield  {author} {\bibinfo {author} {\bibfnamefont {X.}~\bibnamefont {Zheng}}, \bibinfo {author} {\bibfnamefont {A.}~\bibnamefont {Cal{\`{o}}}}, \bibinfo {author} {\bibfnamefont {E.}~\bibnamefont {Albisetti}}, \bibinfo {author} {\bibfnamefont {X.~X.}\ \bibnamefont {Liu}}, \bibinfo {author} {\bibfnamefont {A.~S.~M.}\ \bibnamefont {Alharbi}}, \bibinfo {author} {\bibfnamefont {G.}~\bibnamefont {Arefe}}, \bibinfo {author} {\bibfnamefont {X.~X.}\ \bibnamefont {Liu}}, \bibinfo {author} {\bibfnamefont {M.}~\bibnamefont {Spieser}}, \bibinfo {author} {\bibfnamefont {W.~J.}\ \bibnamefont {Yoo}}, \bibinfo {author} {\bibfnamefont {T.}~\bibnamefont {Taniguchi}}, \bibinfo {author} {\bibfnamefont {K.}~\bibnamefont {Watanabe}}, \bibinfo {author} {\bibfnamefont {C.}~\bibnamefont {Aruta}}, \bibinfo {author} {\bibfnamefont {A.}~\bibnamefont {Ciarrocchi}}, \bibinfo {author} {\bibfnamefont {A.}~\bibnamefont {Kis}}, \bibinfo {author} {\bibfnamefont {B.~S.}\ \bibnamefont {Lee}}, \bibinfo {author} {\bibfnamefont
  {M.}~\bibnamefont {Lipson}}, \bibinfo {author} {\bibfnamefont {J.}~\bibnamefont {Hone}}, \bibinfo {author} {\bibfnamefont {D.}~\bibnamefont {Shahrjerdi}},\ and\ \bibinfo {author} {\bibfnamefont {E.}~\bibnamefont {Riedo}},\ }\bibfield  {title} {\bibinfo {title} {{Patterning metal contacts on monolayer MoS$_2$ with vanishing Schottky barriers using thermal nanolithography}},\ }\href@noop {} {\bibfield  {journal} {\bibinfo  {journal} {Nature Electronics}\ }\textbf {\bibinfo {volume} {2}},\ \bibinfo {pages} {17} (\bibinfo {year} {2019}{\natexlab{a}})}\BibitemShut {NoStop}%
\bibitem [{\citenamefont {Albisetti}\ \emph {et~al.}(2022)\citenamefont {Albisetti}, \citenamefont {Cal{\`{o}}}, \citenamefont {Zanut}, \citenamefont {Zheng}, \citenamefont {de~Peppo},\ and\ \citenamefont {Riedo}}]{Albisetti2022}%
  \BibitemOpen
  \bibfield  {author} {\bibinfo {author} {\bibfnamefont {E.}~\bibnamefont {Albisetti}}, \bibinfo {author} {\bibfnamefont {A.}~\bibnamefont {Cal{\`{o}}}}, \bibinfo {author} {\bibfnamefont {A.}~\bibnamefont {Zanut}}, \bibinfo {author} {\bibfnamefont {X.}~\bibnamefont {Zheng}}, \bibinfo {author} {\bibfnamefont {G.~M.}\ \bibnamefont {de~Peppo}},\ and\ \bibinfo {author} {\bibfnamefont {E.}~\bibnamefont {Riedo}},\ }\bibfield  {title} {\bibinfo {title} {{Thermal scanning probe lithography}},\ }\href@noop {} {\bibfield  {journal} {\bibinfo  {journal} {Nature Reviews Methods Primers}\ }\textbf {\bibinfo {volume} {2}},\ \bibinfo {pages} {32} (\bibinfo {year} {2022})}\BibitemShut {NoStop}%
\bibitem [{\citenamefont {Pirkle}\ \emph {et~al.}(2011)\citenamefont {Pirkle}, \citenamefont {Chan}, \citenamefont {Venugopal}, \citenamefont {Hinojos}, \citenamefont {Magnuson}, \citenamefont {McDonnell}, \citenamefont {Colombo}, \citenamefont {Vogel}, \citenamefont {Ruoff},\ and\ \citenamefont {Wallace}}]{Pirkle2011}%
  \BibitemOpen
  \bibfield  {author} {\bibinfo {author} {\bibfnamefont {A.}~\bibnamefont {Pirkle}}, \bibinfo {author} {\bibfnamefont {J.}~\bibnamefont {Chan}}, \bibinfo {author} {\bibfnamefont {A.}~\bibnamefont {Venugopal}}, \bibinfo {author} {\bibfnamefont {D.}~\bibnamefont {Hinojos}}, \bibinfo {author} {\bibfnamefont {C.~W.}\ \bibnamefont {Magnuson}}, \bibinfo {author} {\bibfnamefont {S.}~\bibnamefont {McDonnell}}, \bibinfo {author} {\bibfnamefont {L.}~\bibnamefont {Colombo}}, \bibinfo {author} {\bibfnamefont {E.~M.}\ \bibnamefont {Vogel}}, \bibinfo {author} {\bibfnamefont {R.~S.}\ \bibnamefont {Ruoff}},\ and\ \bibinfo {author} {\bibfnamefont {R.~M.}\ \bibnamefont {Wallace}},\ }\bibfield  {title} {\bibinfo {title} {{The effect of chemical residues on the physical and electrical properties of chemical vapor deposited graphene transferred to SiO$_2$}},\ }\href@noop {} {\bibfield  {journal} {\bibinfo  {journal} {Applied Physics Letters}\ }\textbf {\bibinfo {volume} {99}},\ \bibinfo {pages} {122108} (\bibinfo {year}
  {2011})}\BibitemShut {NoStop}%
\bibitem [{\citenamefont {Suk}\ \emph {et~al.}(2013)\citenamefont {Suk}, \citenamefont {Lee}, \citenamefont {Lee}, \citenamefont {Chou}, \citenamefont {Piner}, \citenamefont {Hao}, \citenamefont {Akinwande},\ and\ \citenamefont {Ruoff}}]{Suk2013}%
  \BibitemOpen
  \bibfield  {author} {\bibinfo {author} {\bibfnamefont {J.~W.}\ \bibnamefont {Suk}}, \bibinfo {author} {\bibfnamefont {W.~H.}\ \bibnamefont {Lee}}, \bibinfo {author} {\bibfnamefont {J.}~\bibnamefont {Lee}}, \bibinfo {author} {\bibfnamefont {H.}~\bibnamefont {Chou}}, \bibinfo {author} {\bibfnamefont {R.~D.}\ \bibnamefont {Piner}}, \bibinfo {author} {\bibfnamefont {Y.}~\bibnamefont {Hao}}, \bibinfo {author} {\bibfnamefont {D.}~\bibnamefont {Akinwande}},\ and\ \bibinfo {author} {\bibfnamefont {R.~S.}\ \bibnamefont {Ruoff}},\ }\bibfield  {title} {\bibinfo {title} {Enhancement of the electrical properties of graphene grown by chemical vapor deposition via controlling the effects of polymer residue},\ }\href@noop {} {\bibfield  {journal} {\bibinfo  {journal} {Nano Letters}\ }\textbf {\bibinfo {volume} {13}},\ \bibinfo {pages} {1462} (\bibinfo {year} {2013})}\BibitemShut {NoStop}%
\bibitem [{\citenamefont {Lau}\ \emph {et~al.}(2022)\citenamefont {Lau}, \citenamefont {Chee}, \citenamefont {Cao}, \citenamefont {Ooi}, \citenamefont {Tong}, \citenamefont {Bosman}, \citenamefont {Bussolotti}, \citenamefont {Deng}, \citenamefont {Wu}, \citenamefont {Yang}, \citenamefont {Wang}, \citenamefont {Teo}, \citenamefont {Wong}, \citenamefont {Chai}, \citenamefont {Chen}, \citenamefont {Zhang}, \citenamefont {Ang}, \citenamefont {Ang},\ and\ \citenamefont {Goh}}]{Lau2021}%
  \BibitemOpen
  \bibfield  {author} {\bibinfo {author} {\bibfnamefont {C.~S.}\ \bibnamefont {Lau}}, \bibinfo {author} {\bibfnamefont {J.~Y.}\ \bibnamefont {Chee}}, \bibinfo {author} {\bibfnamefont {L.}~\bibnamefont {Cao}}, \bibinfo {author} {\bibfnamefont {Z.}~\bibnamefont {Ooi}}, \bibinfo {author} {\bibfnamefont {S.~W.}\ \bibnamefont {Tong}}, \bibinfo {author} {\bibfnamefont {M.}~\bibnamefont {Bosman}}, \bibinfo {author} {\bibfnamefont {F.}~\bibnamefont {Bussolotti}}, \bibinfo {author} {\bibfnamefont {T.}~\bibnamefont {Deng}}, \bibinfo {author} {\bibfnamefont {G.}~\bibnamefont {Wu}}, \bibinfo {author} {\bibfnamefont {S.-W.}\ \bibnamefont {Yang}}, \bibinfo {author} {\bibfnamefont {T.}~\bibnamefont {Wang}}, \bibinfo {author} {\bibfnamefont {S.~L.}\ \bibnamefont {Teo}}, \bibinfo {author} {\bibfnamefont {C.~P.~Y.}\ \bibnamefont {Wong}}, \bibinfo {author} {\bibfnamefont {J.~W.}\ \bibnamefont {Chai}}, \bibinfo {author} {\bibfnamefont {L.}~\bibnamefont {Chen}}, \bibinfo {author} {\bibfnamefont {Z.~M.}\ \bibnamefont {Zhang}},
  \bibinfo {author} {\bibfnamefont {K.}~\bibnamefont {Ang}}, \bibinfo {author} {\bibfnamefont {Y.~S.}\ \bibnamefont {Ang}},\ and\ \bibinfo {author} {\bibfnamefont {K.~E.~J.}\ \bibnamefont {Goh}},\ }\bibfield  {title} {\bibinfo {title} {{Gate‐Defined Quantum Confinement in CVD 2D WS$_2$}},\ }\href@noop {} {\bibfield  {journal} {\bibinfo  {journal} {Advanced Materials}\ }\textbf {\bibinfo {volume} {34}},\ \bibinfo {pages} {2103907} (\bibinfo {year} {2022})}\BibitemShut {NoStop}%
\bibitem [{\citenamefont {Kieczka}\ \emph {et~al.}(2023)\citenamefont {Kieczka}, \citenamefont {Durrant}, \citenamefont {Milton}, \citenamefont {Goh}, \citenamefont {Bosman},\ and\ \citenamefont {Shluger}}]{Kieczka2023}%
  \BibitemOpen
  \bibfield  {author} {\bibinfo {author} {\bibfnamefont {D.}~\bibnamefont {Kieczka}}, \bibinfo {author} {\bibfnamefont {T.}~\bibnamefont {Durrant}}, \bibinfo {author} {\bibfnamefont {K.}~\bibnamefont {Milton}}, \bibinfo {author} {\bibfnamefont {K.~E.~J.}\ \bibnamefont {Goh}}, \bibinfo {author} {\bibfnamefont {M.}~\bibnamefont {Bosman}},\ and\ \bibinfo {author} {\bibfnamefont {A.}~\bibnamefont {Shluger}},\ }\bibfield  {title} {\bibinfo {title} {{Defects in WS 2 monolayer calculated with a nonlocal functional: any difference from GGA?}},\ }\href@noop {} {\bibfield  {journal} {\bibinfo  {journal} {Electronic Structure}\ }\textbf {\bibinfo {volume} {5}},\ \bibinfo {pages} {024001} (\bibinfo {year} {2023})}\BibitemShut {NoStop}%
\bibitem [{\citenamefont {Zhang}\ \emph {et~al.}(2009)\citenamefont {Zhang}, \citenamefont {Brar}, \citenamefont {Girit}, \citenamefont {Zettl},\ and\ \citenamefont {Crommie}}]{zhang2009}%
  \BibitemOpen
  \bibfield  {author} {\bibinfo {author} {\bibfnamefont {Y.}~\bibnamefont {Zhang}}, \bibinfo {author} {\bibfnamefont {V.~W.}\ \bibnamefont {Brar}}, \bibinfo {author} {\bibfnamefont {C.}~\bibnamefont {Girit}}, \bibinfo {author} {\bibfnamefont {A.}~\bibnamefont {Zettl}},\ and\ \bibinfo {author} {\bibfnamefont {M.~F.}\ \bibnamefont {Crommie}},\ }\bibfield  {title} {\bibinfo {title} {{Origin of spatial charge inhomogeneity in graphene}},\ }\href@noop {} {\bibfield  {journal} {\bibinfo  {journal} {Nature Physics}\ }\textbf {\bibinfo {volume} {5}},\ \bibinfo {pages} {722} (\bibinfo {year} {2009})}\BibitemShut {NoStop}%
\bibitem [{\citenamefont {Martin}\ \emph {et~al.}(2008)\citenamefont {Martin}, \citenamefont {Akerman}, \citenamefont {Ulbricht}, \citenamefont {Lohmann}, \citenamefont {Smet}, \citenamefont {von Klitzing},\ and\ \citenamefont {Yacoby}}]{Martin2008a}%
  \BibitemOpen
  \bibfield  {author} {\bibinfo {author} {\bibfnamefont {J.}~\bibnamefont {Martin}}, \bibinfo {author} {\bibfnamefont {N.}~\bibnamefont {Akerman}}, \bibinfo {author} {\bibfnamefont {G.}~\bibnamefont {Ulbricht}}, \bibinfo {author} {\bibfnamefont {T.}~\bibnamefont {Lohmann}}, \bibinfo {author} {\bibfnamefont {J.~H.}\ \bibnamefont {Smet}}, \bibinfo {author} {\bibfnamefont {K.}~\bibnamefont {von Klitzing}},\ and\ \bibinfo {author} {\bibfnamefont {A.}~\bibnamefont {Yacoby}},\ }\bibfield  {title} {\bibinfo {title} {{Observation of electron–hole puddles in graphene using a scanning single-electron transistor}},\ }\href@noop {} {\bibfield  {journal} {\bibinfo  {journal} {Nature Physics}\ }\textbf {\bibinfo {volume} {4}},\ \bibinfo {pages} {144} (\bibinfo {year} {2008})}\BibitemShut {NoStop}%
\bibitem [{\citenamefont {Goh}\ \emph {et~al.}(2020)\citenamefont {Goh}, \citenamefont {Bussolotti}, \citenamefont {Lau}, \citenamefont {Kotekar‐Patil}, \citenamefont {Ooi},\ and\ \citenamefont {Chee}}]{Goh2020}%
  \BibitemOpen
  \bibfield  {author} {\bibinfo {author} {\bibfnamefont {K.~E.~J.}\ \bibnamefont {Goh}}, \bibinfo {author} {\bibfnamefont {F.}~\bibnamefont {Bussolotti}}, \bibinfo {author} {\bibfnamefont {C.~S.}\ \bibnamefont {Lau}}, \bibinfo {author} {\bibfnamefont {D.}~\bibnamefont {Kotekar‐Patil}}, \bibinfo {author} {\bibfnamefont {Z.~E.}\ \bibnamefont {Ooi}},\ and\ \bibinfo {author} {\bibfnamefont {J.~Y.}\ \bibnamefont {Chee}},\ }\bibfield  {title} {\bibinfo {title} {{Toward Valley‐Coupled Spin Qubits}},\ }\href@noop {} {\bibfield  {journal} {\bibinfo  {journal} {Advanced Quantum Technologies}\ }\textbf {\bibinfo {volume} {3}},\ \bibinfo {pages} {1900123} (\bibinfo {year} {2020})}\BibitemShut {NoStop}%
\bibitem [{\citenamefont {Liu}\ and\ \citenamefont {Hersam}(2019)}]{Liu2019a}%
  \BibitemOpen
  \bibfield  {author} {\bibinfo {author} {\bibfnamefont {X.}~\bibnamefont {Liu}}\ and\ \bibinfo {author} {\bibfnamefont {M.~C.}\ \bibnamefont {Hersam}},\ }\bibfield  {title} {\bibinfo {title} {{2D materials for quantum information science}},\ }\href@noop {} {\bibfield  {journal} {\bibinfo  {journal} {Nature Reviews Materials}\ }\textbf {\bibinfo {volume} {4}},\ \bibinfo {pages} {669} (\bibinfo {year} {2019})}\BibitemShut {NoStop}%
\bibitem [{\citenamefont {Hanson}\ \emph {et~al.}(2007)\citenamefont {Hanson}, \citenamefont {Kouwenhoven}, \citenamefont {Petta}, \citenamefont {Tarucha},\ and\ \citenamefont {Vandersypen}}]{Hanson2007}%
  \BibitemOpen
  \bibfield  {author} {\bibinfo {author} {\bibfnamefont {R.}~\bibnamefont {Hanson}}, \bibinfo {author} {\bibfnamefont {L.~P.}\ \bibnamefont {Kouwenhoven}}, \bibinfo {author} {\bibfnamefont {J.~R.}\ \bibnamefont {Petta}}, \bibinfo {author} {\bibfnamefont {S.}~\bibnamefont {Tarucha}},\ and\ \bibinfo {author} {\bibfnamefont {L.~M.~K.}\ \bibnamefont {Vandersypen}},\ }\bibfield  {title} {\bibinfo {title} {{Spins in few-electron quantum dots}},\ }\href@noop {} {\bibfield  {journal} {\bibinfo  {journal} {Reviews of Modern Physics}\ }\textbf {\bibinfo {volume} {79}},\ \bibinfo {pages} {1217} (\bibinfo {year} {2007})}\BibitemShut {NoStop}%
\bibitem [{\citenamefont {Montblanch}\ \emph {et~al.}(2023)\citenamefont {Montblanch}, \citenamefont {Barbone}, \citenamefont {Aharonovich}, \citenamefont {Atat{\"{u}}re},\ and\ \citenamefont {Ferrari}}]{Montblanch2023}%
  \BibitemOpen
  \bibfield  {author} {\bibinfo {author} {\bibfnamefont {A.~R.}\ \bibnamefont {Montblanch}}, \bibinfo {author} {\bibfnamefont {M.}~\bibnamefont {Barbone}}, \bibinfo {author} {\bibfnamefont {I.}~\bibnamefont {Aharonovich}}, \bibinfo {author} {\bibfnamefont {M.}~\bibnamefont {Atat{\"{u}}re}},\ and\ \bibinfo {author} {\bibfnamefont {A.~C.}\ \bibnamefont {Ferrari}},\ }\bibfield  {title} {\bibinfo {title} {{Layered materials as a platform for quantum technologies}},\ }\href@noop {} {\bibfield  {journal} {\bibinfo  {journal} {Nature Nanotechnology}\ }\textbf {\bibinfo {volume} {18}},\ \bibinfo {pages} {555} (\bibinfo {year} {2023})}\BibitemShut {NoStop}%
\bibitem [{\citenamefont {de~Leon}\ \emph {et~al.}(2021)\citenamefont {de~Leon}, \citenamefont {Itoh}, \citenamefont {Kim}, \citenamefont {Mehta}, \citenamefont {Northup}, \citenamefont {Paik}, \citenamefont {Palmer}, \citenamefont {Samarth}, \citenamefont {Sangtawesin},\ and\ \citenamefont {Steuerman}}]{DeLeon2021a}%
  \BibitemOpen
  \bibfield  {author} {\bibinfo {author} {\bibfnamefont {N.~P.}\ \bibnamefont {de~Leon}}, \bibinfo {author} {\bibfnamefont {K.~M.}\ \bibnamefont {Itoh}}, \bibinfo {author} {\bibfnamefont {D.}~\bibnamefont {Kim}}, \bibinfo {author} {\bibfnamefont {K.~K.}\ \bibnamefont {Mehta}}, \bibinfo {author} {\bibfnamefont {T.~E.}\ \bibnamefont {Northup}}, \bibinfo {author} {\bibfnamefont {H.}~\bibnamefont {Paik}}, \bibinfo {author} {\bibfnamefont {B.~S.}\ \bibnamefont {Palmer}}, \bibinfo {author} {\bibfnamefont {N.}~\bibnamefont {Samarth}}, \bibinfo {author} {\bibfnamefont {S.}~\bibnamefont {Sangtawesin}},\ and\ \bibinfo {author} {\bibfnamefont {D.~W.}\ \bibnamefont {Steuerman}},\ }\bibfield  {title} {\bibinfo {title} {{Materials challenges and opportunities for quantum computing hardware}},\ }\href@noop {} {\bibfield  {journal} {\bibinfo  {journal} {Science}\ }\textbf {\bibinfo {volume} {372}},\ \bibinfo {pages} {eabb2823} (\bibinfo {year} {2021})}\BibitemShut {NoStop}%
\bibitem [{\citenamefont {Lau}\ \emph {et~al.}(2019)\citenamefont {Lau}, \citenamefont {Chee}, \citenamefont {Thian}, \citenamefont {Kawai}, \citenamefont {Deng}, \citenamefont {Wong}, \citenamefont {Ooi}, \citenamefont {Lim},\ and\ \citenamefont {Goh}}]{Lau2019a}%
  \BibitemOpen
  \bibfield  {author} {\bibinfo {author} {\bibfnamefont {C.~S.}\ \bibnamefont {Lau}}, \bibinfo {author} {\bibfnamefont {J.~Y.}\ \bibnamefont {Chee}}, \bibinfo {author} {\bibfnamefont {D.}~\bibnamefont {Thian}}, \bibinfo {author} {\bibfnamefont {H.}~\bibnamefont {Kawai}}, \bibinfo {author} {\bibfnamefont {J.}~\bibnamefont {Deng}}, \bibinfo {author} {\bibfnamefont {S.~L.}\ \bibnamefont {Wong}}, \bibinfo {author} {\bibfnamefont {Z.~E.}\ \bibnamefont {Ooi}}, \bibinfo {author} {\bibfnamefont {Y.-F.}\ \bibnamefont {Lim}},\ and\ \bibinfo {author} {\bibfnamefont {K.~E.~J.}\ \bibnamefont {Goh}},\ }\bibfield  {title} {\bibinfo {title} {{Carrier control in 2D transition metal dichalcogenides with Al$_2$O$_3$ dielectric}},\ }\href@noop {} {\bibfield  {journal} {\bibinfo  {journal} {Scientific Reports}\ }\textbf {\bibinfo {volume} {9}},\ \bibinfo {pages} {8769} (\bibinfo {year} {2019})}\BibitemShut {NoStop}%
\bibitem [{\citenamefont {Zhang}\ \emph {et~al.}(2017)\citenamefont {Zhang}, \citenamefont {Song}, \citenamefont {Luo}, \citenamefont {Deng}, \citenamefont {Mosallanejad}, \citenamefont {Taniguchi}, \citenamefont {Watanabe}, \citenamefont {Li}, \citenamefont {Cao}, \citenamefont {Guo}, \citenamefont {Nori},\ and\ \citenamefont {Guo}}]{Zhang2017i}%
  \BibitemOpen
  \bibfield  {author} {\bibinfo {author} {\bibfnamefont {Z.-Z.~Z.}\ \bibnamefont {Zhang}}, \bibinfo {author} {\bibfnamefont {X.-X.~X.}\ \bibnamefont {Song}}, \bibinfo {author} {\bibfnamefont {G.}~\bibnamefont {Luo}}, \bibinfo {author} {\bibfnamefont {G.-W.~W.}\ \bibnamefont {Deng}}, \bibinfo {author} {\bibfnamefont {V.}~\bibnamefont {Mosallanejad}}, \bibinfo {author} {\bibfnamefont {T.}~\bibnamefont {Taniguchi}}, \bibinfo {author} {\bibfnamefont {K.}~\bibnamefont {Watanabe}}, \bibinfo {author} {\bibfnamefont {H.-O.~O.}\ \bibnamefont {Li}}, \bibinfo {author} {\bibfnamefont {G.}~\bibnamefont {Cao}}, \bibinfo {author} {\bibfnamefont {G.-C. C. G.-P.~P.}\ \bibnamefont {Guo}}, \bibinfo {author} {\bibfnamefont {F.}~\bibnamefont {Nori}},\ and\ \bibinfo {author} {\bibfnamefont {G.-C. C. G.-P.~P.}\ \bibnamefont {Guo}},\ }\bibfield  {title} {\bibinfo {title} {{Electrotunable artificial molecules based on van der Waals heterostructures}},\ }\href@noop {} {\bibfield  {journal} {\bibinfo  {journal} {Science Advances}\
  }\textbf {\bibinfo {volume} {3}},\ \bibinfo {pages} {1} (\bibinfo {year} {2017})}\BibitemShut {NoStop}%
\bibitem [{\citenamefont {Hamer}\ \emph {et~al.}(2018)\citenamefont {Hamer}, \citenamefont {T{\'{o}}v{\'{a}}ri}, \citenamefont {Zhu}, \citenamefont {Thompson}, \citenamefont {Mayorov}, \citenamefont {Prance}, \citenamefont {Lee}, \citenamefont {Haley}, \citenamefont {Kudrynskyi}, \citenamefont {Patan{\`{e}}}, \citenamefont {Terry}, \citenamefont {Kovalyuk}, \citenamefont {Ensslin}, \citenamefont {Kretinin}, \citenamefont {Geim},\ and\ \citenamefont {Gorbachev}}]{Hamer2018}%
  \BibitemOpen
  \bibfield  {author} {\bibinfo {author} {\bibfnamefont {M.}~\bibnamefont {Hamer}}, \bibinfo {author} {\bibfnamefont {E.}~\bibnamefont {T{\'{o}}v{\'{a}}ri}}, \bibinfo {author} {\bibfnamefont {M.}~\bibnamefont {Zhu}}, \bibinfo {author} {\bibfnamefont {M.~D.}\ \bibnamefont {Thompson}}, \bibinfo {author} {\bibfnamefont {A.}~\bibnamefont {Mayorov}}, \bibinfo {author} {\bibfnamefont {J.}~\bibnamefont {Prance}}, \bibinfo {author} {\bibfnamefont {Y.}~\bibnamefont {Lee}}, \bibinfo {author} {\bibfnamefont {R.~P.}\ \bibnamefont {Haley}}, \bibinfo {author} {\bibfnamefont {Z.~R.}\ \bibnamefont {Kudrynskyi}}, \bibinfo {author} {\bibfnamefont {A.}~\bibnamefont {Patan{\`{e}}}}, \bibinfo {author} {\bibfnamefont {D.}~\bibnamefont {Terry}}, \bibinfo {author} {\bibfnamefont {Z.~D.}\ \bibnamefont {Kovalyuk}}, \bibinfo {author} {\bibfnamefont {K.}~\bibnamefont {Ensslin}}, \bibinfo {author} {\bibfnamefont {A.~V.}\ \bibnamefont {Kretinin}}, \bibinfo {author} {\bibfnamefont {A.}~\bibnamefont {Geim}},\ and\ \bibinfo {author}
  {\bibfnamefont {R.}~\bibnamefont {Gorbachev}},\ }\bibfield  {title} {\bibinfo {title} {{Gate-Defined Quantum Confinement in InSe-Based van der Waals Heterostructures}},\ }\href@noop {} {\bibfield  {journal} {\bibinfo  {journal} {Nano Letters}\ }\textbf {\bibinfo {volume} {18}},\ \bibinfo {pages} {3950} (\bibinfo {year} {2018})}\BibitemShut {NoStop}%
\bibitem [{\citenamefont {Wang}\ \emph {et~al.}(2018)\citenamefont {Wang}, \citenamefont {{De Greve}}, \citenamefont {Jauregui}, \citenamefont {Sushko}, \citenamefont {High}, \citenamefont {Zhou}, \citenamefont {Scuri}, \citenamefont {Taniguchi}, \citenamefont {Watanabe}, \citenamefont {Lukin}, \citenamefont {Park},\ and\ \citenamefont {Kim}}]{Wang2018}%
  \BibitemOpen
  \bibfield  {author} {\bibinfo {author} {\bibfnamefont {K.}~\bibnamefont {Wang}}, \bibinfo {author} {\bibfnamefont {K.}~\bibnamefont {{De Greve}}}, \bibinfo {author} {\bibfnamefont {L.~A.}\ \bibnamefont {Jauregui}}, \bibinfo {author} {\bibfnamefont {A.}~\bibnamefont {Sushko}}, \bibinfo {author} {\bibfnamefont {A.}~\bibnamefont {High}}, \bibinfo {author} {\bibfnamefont {Y.}~\bibnamefont {Zhou}}, \bibinfo {author} {\bibfnamefont {G.}~\bibnamefont {Scuri}}, \bibinfo {author} {\bibfnamefont {T.}~\bibnamefont {Taniguchi}}, \bibinfo {author} {\bibfnamefont {K.}~\bibnamefont {Watanabe}}, \bibinfo {author} {\bibfnamefont {M.~D.}\ \bibnamefont {Lukin}}, \bibinfo {author} {\bibfnamefont {H.}~\bibnamefont {Park}},\ and\ \bibinfo {author} {\bibfnamefont {P.}~\bibnamefont {Kim}},\ }\bibfield  {title} {\bibinfo {title} {{Electrical control of charged carriers and excitons in atomically thin materials}},\ }\href@noop {} {\bibfield  {journal} {\bibinfo  {journal} {Nature Nanotechnology}\ }\textbf {\bibinfo {volume} {13}},\
  \bibinfo {pages} {128} (\bibinfo {year} {2018})}\BibitemShut {NoStop}%
\bibitem [{\citenamefont {Pisoni}\ \emph {et~al.}(2018)\citenamefont {Pisoni}, \citenamefont {Lei}, \citenamefont {Back}, \citenamefont {Eich}, \citenamefont {Overweg}, \citenamefont {Lee}, \citenamefont {Watanabe}, \citenamefont {Taniguchi}, \citenamefont {Ihn},\ and\ \citenamefont {Ensslin}}]{Pisoni2018}%
  \BibitemOpen
  \bibfield  {author} {\bibinfo {author} {\bibfnamefont {R.}~\bibnamefont {Pisoni}}, \bibinfo {author} {\bibfnamefont {Z.}~\bibnamefont {Lei}}, \bibinfo {author} {\bibfnamefont {P.}~\bibnamefont {Back}}, \bibinfo {author} {\bibfnamefont {M.}~\bibnamefont {Eich}}, \bibinfo {author} {\bibfnamefont {H.}~\bibnamefont {Overweg}}, \bibinfo {author} {\bibfnamefont {Y.}~\bibnamefont {Lee}}, \bibinfo {author} {\bibfnamefont {K.}~\bibnamefont {Watanabe}}, \bibinfo {author} {\bibfnamefont {T.}~\bibnamefont {Taniguchi}}, \bibinfo {author} {\bibfnamefont {T.}~\bibnamefont {Ihn}},\ and\ \bibinfo {author} {\bibfnamefont {K.}~\bibnamefont {Ensslin}},\ }\bibfield  {title} {\bibinfo {title} {{Gate-Tunable Quantum Dot in a High Quality Single Layer MoS$_2$ Van der Waals Heterostructure}},\ }\href@noop {} {\bibfield  {journal} {\bibinfo  {journal} {Applied Physics Letters}\ }\textbf {\bibinfo {volume} {123101}},\ \bibinfo {pages} {2016} (\bibinfo {year} {2018})}\BibitemShut {NoStop}%
\bibitem [{\citenamefont {Eich}\ \emph {et~al.}(2018)\citenamefont {Eich}, \citenamefont {Pisoni}, \citenamefont {Overweg}, \citenamefont {Kurzmann}, \citenamefont {Lee}, \citenamefont {Rickhaus}, \citenamefont {Ihn}, \citenamefont {Ensslin}, \citenamefont {Herman}, \citenamefont {Sigrist}, \citenamefont {Watanabe},\ and\ \citenamefont {Taniguchi}}]{Eich2018}%
  \BibitemOpen
  \bibfield  {author} {\bibinfo {author} {\bibfnamefont {M.}~\bibnamefont {Eich}}, \bibinfo {author} {\bibfnamefont {R.}~\bibnamefont {Pisoni}}, \bibinfo {author} {\bibfnamefont {H.}~\bibnamefont {Overweg}}, \bibinfo {author} {\bibfnamefont {A.}~\bibnamefont {Kurzmann}}, \bibinfo {author} {\bibfnamefont {Y.}~\bibnamefont {Lee}}, \bibinfo {author} {\bibfnamefont {P.}~\bibnamefont {Rickhaus}}, \bibinfo {author} {\bibfnamefont {T.}~\bibnamefont {Ihn}}, \bibinfo {author} {\bibfnamefont {K.}~\bibnamefont {Ensslin}}, \bibinfo {author} {\bibfnamefont {F.}~\bibnamefont {Herman}}, \bibinfo {author} {\bibfnamefont {M.}~\bibnamefont {Sigrist}}, \bibinfo {author} {\bibfnamefont {K.}~\bibnamefont {Watanabe}},\ and\ \bibinfo {author} {\bibfnamefont {T.}~\bibnamefont {Taniguchi}},\ }\bibfield  {title} {\bibinfo {title} {{Spin and Valley States in Gate-Defined Bilayer Graphene Quantum Dots}},\ }\href {https://doi.org/10.1103/PhysRevX.8.031023} {\bibfield  {journal} {\bibinfo  {journal} {Physical Review X}\ }\textbf {\bibinfo
  {volume} {8}},\ \bibinfo {pages} {1} (\bibinfo {year} {2018})}\BibitemShut {NoStop}%
\bibitem [{\citenamefont {Davari}\ \emph {et~al.}(2020)\citenamefont {Davari}, \citenamefont {Stacy}, \citenamefont {Mercado}, \citenamefont {Tull}, \citenamefont {Basnet}, \citenamefont {Pandey}, \citenamefont {Watanabe}, \citenamefont {Taniguchi}, \citenamefont {Hu},\ and\ \citenamefont {Churchill}}]{Davari2020}%
  \BibitemOpen
  \bibfield  {author} {\bibinfo {author} {\bibfnamefont {S.}~\bibnamefont {Davari}}, \bibinfo {author} {\bibfnamefont {J.}~\bibnamefont {Stacy}}, \bibinfo {author} {\bibfnamefont {A.~M.}\ \bibnamefont {Mercado}}, \bibinfo {author} {\bibfnamefont {J.~D.}\ \bibnamefont {Tull}}, \bibinfo {author} {\bibfnamefont {R.}~\bibnamefont {Basnet}}, \bibinfo {author} {\bibfnamefont {K.}~\bibnamefont {Pandey}}, \bibinfo {author} {\bibfnamefont {K.}~\bibnamefont {Watanabe}}, \bibinfo {author} {\bibfnamefont {T.}~\bibnamefont {Taniguchi}}, \bibinfo {author} {\bibfnamefont {J.}~\bibnamefont {Hu}},\ and\ \bibinfo {author} {\bibfnamefont {H.~O.}\ \bibnamefont {Churchill}},\ }\bibfield  {title} {\bibinfo {title} {{Gate-Defined Accumulation-Mode Quantum Dots in Monolayer and Bilayer WSe$_2$}},\ }\href@noop {} {\bibfield  {journal} {\bibinfo  {journal} {Physical Review Applied}\ }\textbf {\bibinfo {volume} {13}},\ \bibinfo {pages} {1} (\bibinfo {year} {2020})}\BibitemShut {NoStop}%
\bibitem [{\citenamefont {Boddison-Chouinard}\ \emph {et~al.}(2021)\citenamefont {Boddison-Chouinard}, \citenamefont {Bogan}, \citenamefont {Fong}, \citenamefont {Watanabe}, \citenamefont {Taniguchi}, \citenamefont {Studenikin}, \citenamefont {Sachrajda}, \citenamefont {Korkusinski}, \citenamefont {Altintas}, \citenamefont {Bieniek}, \citenamefont {Hawrylak}, \citenamefont {Luican-Mayer},\ and\ \citenamefont {Gaudreau}}]{Boddison-Chouinard2021}%
  \BibitemOpen
  \bibfield  {author} {\bibinfo {author} {\bibfnamefont {J.}~\bibnamefont {Boddison-Chouinard}}, \bibinfo {author} {\bibfnamefont {A.}~\bibnamefont {Bogan}}, \bibinfo {author} {\bibfnamefont {N.}~\bibnamefont {Fong}}, \bibinfo {author} {\bibfnamefont {K.}~\bibnamefont {Watanabe}}, \bibinfo {author} {\bibfnamefont {T.}~\bibnamefont {Taniguchi}}, \bibinfo {author} {\bibfnamefont {S.}~\bibnamefont {Studenikin}}, \bibinfo {author} {\bibfnamefont {A.}~\bibnamefont {Sachrajda}}, \bibinfo {author} {\bibfnamefont {M.}~\bibnamefont {Korkusinski}}, \bibinfo {author} {\bibfnamefont {A.}~\bibnamefont {Altintas}}, \bibinfo {author} {\bibfnamefont {M.}~\bibnamefont {Bieniek}}, \bibinfo {author} {\bibfnamefont {P.}~\bibnamefont {Hawrylak}}, \bibinfo {author} {\bibfnamefont {A.}~\bibnamefont {Luican-Mayer}},\ and\ \bibinfo {author} {\bibfnamefont {L.}~\bibnamefont {Gaudreau}},\ }\bibfield  {title} {\bibinfo {title} {{Gate-controlled quantum dots in monolayer WSe$_2$}},\ }\href@noop {} {\bibfield  {journal} {\bibinfo
  {journal} {Applied Physics Letters}\ }\textbf {\bibinfo {volume} {119}},\ \bibinfo {pages} {133104} (\bibinfo {year} {2021})}\BibitemShut {NoStop}%
\bibitem [{\citenamefont {Wang}\ \emph {et~al.}(2013)\citenamefont {Wang}, \citenamefont {Meric}, \citenamefont {Huang}, \citenamefont {Gao}, \citenamefont {Gao}, \citenamefont {Tran}, \citenamefont {Taniguchi}, \citenamefont {Watanabe}, \citenamefont {Campos}, \citenamefont {Muller}, \citenamefont {Guo}, \citenamefont {Kim}, \citenamefont {Hone}, \citenamefont {Shepard},\ and\ \citenamefont {Dean}}]{Wang2013}%
  \BibitemOpen
  \bibfield  {author} {\bibinfo {author} {\bibfnamefont {L.}~\bibnamefont {Wang}}, \bibinfo {author} {\bibfnamefont {I.}~\bibnamefont {Meric}}, \bibinfo {author} {\bibfnamefont {P.~Y.}\ \bibnamefont {Huang}}, \bibinfo {author} {\bibfnamefont {Q.}~\bibnamefont {Gao}}, \bibinfo {author} {\bibfnamefont {Y.}~\bibnamefont {Gao}}, \bibinfo {author} {\bibfnamefont {H.}~\bibnamefont {Tran}}, \bibinfo {author} {\bibfnamefont {T.}~\bibnamefont {Taniguchi}}, \bibinfo {author} {\bibfnamefont {K.}~\bibnamefont {Watanabe}}, \bibinfo {author} {\bibfnamefont {L.~M.}\ \bibnamefont {Campos}}, \bibinfo {author} {\bibfnamefont {D.~A.}\ \bibnamefont {Muller}}, \bibinfo {author} {\bibfnamefont {J.}~\bibnamefont {Guo}}, \bibinfo {author} {\bibfnamefont {P.}~\bibnamefont {Kim}}, \bibinfo {author} {\bibfnamefont {J.}~\bibnamefont {Hone}}, \bibinfo {author} {\bibfnamefont {K.~L.}\ \bibnamefont {Shepard}},\ and\ \bibinfo {author} {\bibfnamefont {C.~R.}\ \bibnamefont {Dean}},\ }\bibfield  {title} {\bibinfo {title} {One-dimensional
  electrical contact to a two-dimensional material},\ }\href@noop {} {\bibfield  {journal} {\bibinfo  {journal} {Science}\ }\textbf {\bibinfo {volume} {342}},\ \bibinfo {pages} {614} (\bibinfo {year} {2013})}\BibitemShut {NoStop}%
\bibitem [{\citenamefont {Xu}\ \emph {et~al.}(2016)\citenamefont {Xu}, \citenamefont {Wu}, \citenamefont {Lu}, \citenamefont {Han}, \citenamefont {Long}, \citenamefont {Chen}, \citenamefont {Han}, \citenamefont {Ye}, \citenamefont {Wu}, \citenamefont {Lin}, \citenamefont {Shen}, \citenamefont {Cai}, \citenamefont {He}, \citenamefont {Zhang}, \citenamefont {Lortz}, \citenamefont {Cheng},\ and\ \citenamefont {Wang}}]{Xu2016a}%
  \BibitemOpen
  \bibfield  {author} {\bibinfo {author} {\bibfnamefont {S.}~\bibnamefont {Xu}}, \bibinfo {author} {\bibfnamefont {Z.}~\bibnamefont {Wu}}, \bibinfo {author} {\bibfnamefont {H.}~\bibnamefont {Lu}}, \bibinfo {author} {\bibfnamefont {Y.}~\bibnamefont {Han}}, \bibinfo {author} {\bibfnamefont {G.}~\bibnamefont {Long}}, \bibinfo {author} {\bibfnamefont {X.}~\bibnamefont {Chen}}, \bibinfo {author} {\bibfnamefont {T.}~\bibnamefont {Han}}, \bibinfo {author} {\bibfnamefont {W.}~\bibnamefont {Ye}}, \bibinfo {author} {\bibfnamefont {Y.}~\bibnamefont {Wu}}, \bibinfo {author} {\bibfnamefont {J.}~\bibnamefont {Lin}}, \bibinfo {author} {\bibfnamefont {J.}~\bibnamefont {Shen}}, \bibinfo {author} {\bibfnamefont {Y.}~\bibnamefont {Cai}}, \bibinfo {author} {\bibfnamefont {Y.}~\bibnamefont {He}}, \bibinfo {author} {\bibfnamefont {F.}~\bibnamefont {Zhang}}, \bibinfo {author} {\bibfnamefont {R.}~\bibnamefont {Lortz}}, \bibinfo {author} {\bibfnamefont {C.}~\bibnamefont {Cheng}},\ and\ \bibinfo {author} {\bibfnamefont
  {N.}~\bibnamefont {Wang}},\ }\bibfield  {title} {\bibinfo {title} {{Universal low-temperature Ohmic contacts for quantum transport in transition metal dichalcogenides}},\ }\href@noop {} {\bibfield  {journal} {\bibinfo  {journal} {2D Materials}\ }\textbf {\bibinfo {volume} {3}},\ \bibinfo {pages} {021007} (\bibinfo {year} {2016})}\BibitemShut {NoStop}%
\bibitem [{\citenamefont {Lau}\ \emph {et~al.}(2020)\citenamefont {Lau}, \citenamefont {Chee}, \citenamefont {Ang}, \citenamefont {Tong}, \citenamefont {Cao}, \citenamefont {Ooi}, \citenamefont {Wang}, \citenamefont {Ang}, \citenamefont {Wang}, \citenamefont {Chhowalla},\ and\ \citenamefont {Goh}}]{Lau2020}%
  \BibitemOpen
  \bibfield  {author} {\bibinfo {author} {\bibfnamefont {C.~S.}\ \bibnamefont {Lau}}, \bibinfo {author} {\bibfnamefont {J.~Y.}\ \bibnamefont {Chee}}, \bibinfo {author} {\bibfnamefont {Y.~S.}\ \bibnamefont {Ang}}, \bibinfo {author} {\bibfnamefont {S.~W.}\ \bibnamefont {Tong}}, \bibinfo {author} {\bibfnamefont {L.}~\bibnamefont {Cao}}, \bibinfo {author} {\bibfnamefont {Z.-E.}\ \bibnamefont {Ooi}}, \bibinfo {author} {\bibfnamefont {T.}~\bibnamefont {Wang}}, \bibinfo {author} {\bibfnamefont {L.~K.}\ \bibnamefont {Ang}}, \bibinfo {author} {\bibfnamefont {Y.}~\bibnamefont {Wang}}, \bibinfo {author} {\bibfnamefont {M.}~\bibnamefont {Chhowalla}},\ and\ \bibinfo {author} {\bibfnamefont {K.~E.~J.}\ \bibnamefont {Goh}},\ }\bibfield  {title} {\bibinfo {title} {{Quantum Transport in Two-Dimensional WS$_2$ with High-Efficiency Carrier Injection through Indium Alloy Contacts}},\ }\href@noop {} {\bibfield  {journal} {\bibinfo  {journal} {ACS Nano}\ }\textbf {\bibinfo {volume} {14}},\ \bibinfo {pages} {13700} (\bibinfo {year}
  {2020})}\BibitemShut {NoStop}%
\bibitem [{\citenamefont {Zheng}\ \emph {et~al.}(2019{\natexlab{b}})\citenamefont {Zheng}, \citenamefont {Cal{\`{o}}}, \citenamefont {Albisetti}, \citenamefont {Liu}, \citenamefont {Alharbi}, \citenamefont {Arefe}, \citenamefont {Liu}, \citenamefont {Spieser}, \citenamefont {Yoo}, \citenamefont {Taniguchi}, \citenamefont {Watanabe}, \citenamefont {Aruta}, \citenamefont {Ciarrocchi}, \citenamefont {Kis}, \citenamefont {Lee}, \citenamefont {Lipson}, \citenamefont {Hone}, \citenamefont {Shahrjerdi},\ and\ \citenamefont {Riedo}}]{Zheng2019b}%
  \BibitemOpen
  \bibfield  {author} {\bibinfo {author} {\bibfnamefont {X.}~\bibnamefont {Zheng}}, \bibinfo {author} {\bibfnamefont {A.}~\bibnamefont {Cal{\`{o}}}}, \bibinfo {author} {\bibfnamefont {E.}~\bibnamefont {Albisetti}}, \bibinfo {author} {\bibfnamefont {X.}~\bibnamefont {Liu}}, \bibinfo {author} {\bibfnamefont {A.~S.~M.}\ \bibnamefont {Alharbi}}, \bibinfo {author} {\bibfnamefont {G.}~\bibnamefont {Arefe}}, \bibinfo {author} {\bibfnamefont {X.}~\bibnamefont {Liu}}, \bibinfo {author} {\bibfnamefont {M.}~\bibnamefont {Spieser}}, \bibinfo {author} {\bibfnamefont {W.~J.}\ \bibnamefont {Yoo}}, \bibinfo {author} {\bibfnamefont {T.}~\bibnamefont {Taniguchi}}, \bibinfo {author} {\bibfnamefont {K.}~\bibnamefont {Watanabe}}, \bibinfo {author} {\bibfnamefont {C.}~\bibnamefont {Aruta}}, \bibinfo {author} {\bibfnamefont {A.}~\bibnamefont {Ciarrocchi}}, \bibinfo {author} {\bibfnamefont {A.}~\bibnamefont {Kis}}, \bibinfo {author} {\bibfnamefont {B.~S.}\ \bibnamefont {Lee}}, \bibinfo {author} {\bibfnamefont {M.}~\bibnamefont
  {Lipson}}, \bibinfo {author} {\bibfnamefont {J.}~\bibnamefont {Hone}}, \bibinfo {author} {\bibfnamefont {D.}~\bibnamefont {Shahrjerdi}},\ and\ \bibinfo {author} {\bibfnamefont {E.}~\bibnamefont {Riedo}},\ }\bibfield  {title} {\bibinfo {title} {{Patterning metal contacts on monolayer MoS$_2$ with vanishing Schottky barriers using thermal nanolithography}},\ }\href@noop {} {\bibfield  {journal} {\bibinfo  {journal} {Nature Electronics}\ }\textbf {\bibinfo {volume} {2}},\ \bibinfo {pages} {17} (\bibinfo {year} {2019}{\natexlab{b}})}\BibitemShut {NoStop}%
\bibitem [{\citenamefont {Liu}\ \emph {et~al.}(2010)\citenamefont {Liu}, \citenamefont {Hug},\ and\ \citenamefont {Vandersypen}}]{Liu2010}%
  \BibitemOpen
  \bibfield  {author} {\bibinfo {author} {\bibfnamefont {X.~L.}\ \bibnamefont {Liu}}, \bibinfo {author} {\bibfnamefont {D.}~\bibnamefont {Hug}},\ and\ \bibinfo {author} {\bibfnamefont {L.~M.~K.}\ \bibnamefont {Vandersypen}},\ }\bibfield  {title} {\bibinfo {title} {Gate-defined graphene double quantum dot and excited state spectroscopy},\ }\href@noop {} {\bibfield  {journal} {\bibinfo  {journal} {Nano Letters}\ }\textbf {\bibinfo {volume} {10}},\ \bibinfo {pages} {1623} (\bibinfo {year} {2010})}\BibitemShut {NoStop}%
\bibitem [{\citenamefont {Lee}\ \emph {et~al.}(2016)\citenamefont {Lee}, \citenamefont {Kulkarni},\ and\ \citenamefont {Zhong}}]{Lee2016}%
  \BibitemOpen
  \bibfield  {author} {\bibinfo {author} {\bibfnamefont {K.}~\bibnamefont {Lee}}, \bibinfo {author} {\bibfnamefont {G.}~\bibnamefont {Kulkarni}},\ and\ \bibinfo {author} {\bibfnamefont {Z.}~\bibnamefont {Zhong}},\ }\bibfield  {title} {\bibinfo {title} {{Coulomb blockade in monolayer MoS$_2$ single electron transistor}},\ }\href@noop {} {\bibfield  {journal} {\bibinfo  {journal} {Nanoscale}\ }\textbf {\bibinfo {volume} {8}},\ \bibinfo {pages} {7755} (\bibinfo {year} {2016})}\BibitemShut {NoStop}%
\bibitem [{\citenamefont {Brotons-Gisbert}\ \emph {et~al.}(2019)\citenamefont {Brotons-Gisbert}, \citenamefont {Branny}, \citenamefont {Kumar}, \citenamefont {Picard}, \citenamefont {Proux}, \citenamefont {Gray}, \citenamefont {Burch}, \citenamefont {Watanabe}, \citenamefont {Taniguchi},\ and\ \citenamefont {Gerardot}}]{Brotons-Gisbert2019d}%
  \BibitemOpen
  \bibfield  {author} {\bibinfo {author} {\bibfnamefont {M.}~\bibnamefont {Brotons-Gisbert}}, \bibinfo {author} {\bibfnamefont {A.}~\bibnamefont {Branny}}, \bibinfo {author} {\bibfnamefont {S.}~\bibnamefont {Kumar}}, \bibinfo {author} {\bibfnamefont {R.}~\bibnamefont {Picard}}, \bibinfo {author} {\bibfnamefont {R.}~\bibnamefont {Proux}}, \bibinfo {author} {\bibfnamefont {M.}~\bibnamefont {Gray}}, \bibinfo {author} {\bibfnamefont {K.~S.}\ \bibnamefont {Burch}}, \bibinfo {author} {\bibfnamefont {K.}~\bibnamefont {Watanabe}}, \bibinfo {author} {\bibfnamefont {T.}~\bibnamefont {Taniguchi}},\ and\ \bibinfo {author} {\bibfnamefont {B.~D.}\ \bibnamefont {Gerardot}},\ }\bibfield  {title} {\bibinfo {title} {{Coulomb blockade in an atomically thin quantum dot coupled to a tunable Fermi reservoir}},\ }\href@noop {} {\bibfield  {journal} {\bibinfo  {journal} {Nature Nanotechnology}\ }\textbf {\bibinfo {volume} {14}},\ \bibinfo {pages} {442} (\bibinfo {year} {2019})}\BibitemShut {NoStop}%
\bibitem [{\citenamefont {Zhang}\ \emph {et~al.}(2023)\citenamefont {Zhang}, \citenamefont {Venkatakrishnarao}, \citenamefont {Bosman}, \citenamefont {Fu}, \citenamefont {Das}, \citenamefont {Bussolotti}, \citenamefont {Lee}, \citenamefont {Teo}, \citenamefont {Huang}, \citenamefont {Verzhbitskiy}, \citenamefont {Jiang}, \citenamefont {Jiang}, \citenamefont {Chai}, \citenamefont {Tong}, \citenamefont {Ooi}, \citenamefont {Wong}, \citenamefont {Ang}, \citenamefont {Goh},\ and\ \citenamefont {Lau}}]{Zhang2023}%
  \BibitemOpen
  \bibfield  {author} {\bibinfo {author} {\bibfnamefont {Y.}~\bibnamefont {Zhang}}, \bibinfo {author} {\bibfnamefont {D.}~\bibnamefont {Venkatakrishnarao}}, \bibinfo {author} {\bibfnamefont {M.}~\bibnamefont {Bosman}}, \bibinfo {author} {\bibfnamefont {W.}~\bibnamefont {Fu}}, \bibinfo {author} {\bibfnamefont {S.}~\bibnamefont {Das}}, \bibinfo {author} {\bibfnamefont {F.}~\bibnamefont {Bussolotti}}, \bibinfo {author} {\bibfnamefont {R.}~\bibnamefont {Lee}}, \bibinfo {author} {\bibfnamefont {S.~L.}\ \bibnamefont {Teo}}, \bibinfo {author} {\bibfnamefont {D.}~\bibnamefont {Huang}}, \bibinfo {author} {\bibfnamefont {I.}~\bibnamefont {Verzhbitskiy}}, \bibinfo {author} {\bibfnamefont {Z.}~\bibnamefont {Jiang}}, \bibinfo {author} {\bibfnamefont {Z.}~\bibnamefont {Jiang}}, \bibinfo {author} {\bibfnamefont {J.}~\bibnamefont {Chai}}, \bibinfo {author} {\bibfnamefont {S.~W.}\ \bibnamefont {Tong}}, \bibinfo {author} {\bibfnamefont {Z.-E.}\ \bibnamefont {Ooi}}, \bibinfo {author} {\bibfnamefont {C.~P.~Y.}\ \bibnamefont
  {Wong}}, \bibinfo {author} {\bibfnamefont {Y.~S.}\ \bibnamefont {Ang}}, \bibinfo {author} {\bibfnamefont {K.~E.~J.}\ \bibnamefont {Goh}},\ and\ \bibinfo {author} {\bibfnamefont {C.~S.}\ \bibnamefont {Lau}},\ }\bibfield  {title} {\bibinfo {title} {{Liquid-Metal-Printed Ultrathin Oxides for Atomically Smooth 2D Material Heterostructures}},\ }\href@noop {} {\bibfield  {journal} {\bibinfo  {journal} {ACS Nano}\ }\textbf {\bibinfo {volume} {17}},\ \bibinfo {pages} {7929} (\bibinfo {year} {2023})}\BibitemShut {NoStop}%
\bibitem [{\citenamefont {Lau}\ \emph {et~al.}(2023)\citenamefont {Lau}, \citenamefont {Das}, \citenamefont {Verzhbitskiy}, \citenamefont {Huang}, \citenamefont {Zhang}, \citenamefont {Talha-Dean}, \citenamefont {Fu}, \citenamefont {Venkatakrishnarao},\ and\ \citenamefont {Johnson~Goh}}]{Lau2023}%
  \BibitemOpen
  \bibfield  {author} {\bibinfo {author} {\bibfnamefont {C.~S.}\ \bibnamefont {Lau}}, \bibinfo {author} {\bibfnamefont {S.}~\bibnamefont {Das}}, \bibinfo {author} {\bibfnamefont {I.~A.}\ \bibnamefont {Verzhbitskiy}}, \bibinfo {author} {\bibfnamefont {D.}~\bibnamefont {Huang}}, \bibinfo {author} {\bibfnamefont {Y.}~\bibnamefont {Zhang}}, \bibinfo {author} {\bibfnamefont {T.}~\bibnamefont {Talha-Dean}}, \bibinfo {author} {\bibfnamefont {W.}~\bibnamefont {Fu}}, \bibinfo {author} {\bibfnamefont {D.}~\bibnamefont {Venkatakrishnarao}},\ and\ \bibinfo {author} {\bibfnamefont {K.~E.}\ \bibnamefont {Johnson~Goh}},\ }\bibfield  {title} {\bibinfo {title} {Dielectrics for two-dimensional transition-metal dichalcogenide applications},\ }\href@noop {} {\bibfield  {journal} {\bibinfo  {journal} {ACS Nano}\ }\textbf {\bibinfo {volume} {17}},\ \bibinfo {pages} {9870} (\bibinfo {year} {2023})}\BibitemShut {NoStop}%
\bibitem [{\citenamefont {Das}\ \emph {et~al.}(2021)\citenamefont {Das}, \citenamefont {Sebastian}, \citenamefont {Pop}, \citenamefont {McClellan}, \citenamefont {Franklin}, \citenamefont {Grasser}, \citenamefont {Knobloch}, \citenamefont {Illarionov}, \citenamefont {Penumatcha}, \citenamefont {Appenzeller}, \citenamefont {Chen}, \citenamefont {Zhu}, \citenamefont {Asselberghs}, \citenamefont {Li}, \citenamefont {Avci}, \citenamefont {Bhat}, \citenamefont {Anthopoulos},\ and\ \citenamefont {Singh}}]{Das2021}%
  \BibitemOpen
  \bibfield  {author} {\bibinfo {author} {\bibfnamefont {S.}~\bibnamefont {Das}}, \bibinfo {author} {\bibfnamefont {A.}~\bibnamefont {Sebastian}}, \bibinfo {author} {\bibfnamefont {E.}~\bibnamefont {Pop}}, \bibinfo {author} {\bibfnamefont {C.~J.}\ \bibnamefont {McClellan}}, \bibinfo {author} {\bibfnamefont {A.~D.}\ \bibnamefont {Franklin}}, \bibinfo {author} {\bibfnamefont {T.}~\bibnamefont {Grasser}}, \bibinfo {author} {\bibfnamefont {T.}~\bibnamefont {Knobloch}}, \bibinfo {author} {\bibfnamefont {Y.}~\bibnamefont {Illarionov}}, \bibinfo {author} {\bibfnamefont {A.~V.}\ \bibnamefont {Penumatcha}}, \bibinfo {author} {\bibfnamefont {J.}~\bibnamefont {Appenzeller}}, \bibinfo {author} {\bibfnamefont {Z.}~\bibnamefont {Chen}}, \bibinfo {author} {\bibfnamefont {W.}~\bibnamefont {Zhu}}, \bibinfo {author} {\bibfnamefont {I.}~\bibnamefont {Asselberghs}}, \bibinfo {author} {\bibfnamefont {L.~J.}\ \bibnamefont {Li}}, \bibinfo {author} {\bibfnamefont {U.~E.}\ \bibnamefont {Avci}}, \bibinfo {author} {\bibfnamefont
  {N.}~\bibnamefont {Bhat}}, \bibinfo {author} {\bibfnamefont {T.~D.}\ \bibnamefont {Anthopoulos}},\ and\ \bibinfo {author} {\bibfnamefont {R.}~\bibnamefont {Singh}},\ }\bibfield  {title} {\bibinfo {title} {{Transistors based on two-dimensional materials for future integrated circuits}},\ }\href@noop {} {\bibfield  {journal} {\bibinfo  {journal} {Nature Electronics}\ }\textbf {\bibinfo {volume} {4}},\ \bibinfo {pages} {786} (\bibinfo {year} {2021})}\BibitemShut {NoStop}%
\bibitem [{\citenamefont {Tan}\ \emph {et~al.}(2016)\citenamefont {Tan}, \citenamefont {Fan}, \citenamefont {Rong}, \citenamefont {Porter}, \citenamefont {Lau}, \citenamefont {Zhou}, \citenamefont {He}, \citenamefont {Wang}, \citenamefont {Bhaskaran},\ and\ \citenamefont {Warner}}]{Tan2016}%
  \BibitemOpen
  \bibfield  {author} {\bibinfo {author} {\bibfnamefont {H.}~\bibnamefont {Tan}}, \bibinfo {author} {\bibfnamefont {Y.}~\bibnamefont {Fan}}, \bibinfo {author} {\bibfnamefont {Y.}~\bibnamefont {Rong}}, \bibinfo {author} {\bibfnamefont {B.}~\bibnamefont {Porter}}, \bibinfo {author} {\bibfnamefont {C.}~\bibnamefont {Lau}}, \bibinfo {author} {\bibfnamefont {Y.}~\bibnamefont {Zhou}}, \bibinfo {author} {\bibfnamefont {Z.}~\bibnamefont {He}}, \bibinfo {author} {\bibfnamefont {S.}~\bibnamefont {Wang}}, \bibinfo {author} {\bibfnamefont {H.}~\bibnamefont {Bhaskaran}},\ and\ \bibinfo {author} {\bibfnamefont {J.}~\bibnamefont {Warner}},\ }\bibfield  {title} {\bibinfo {title} {{Doping Graphene Transistors Using Vertical Stacked Monolayer WS$_2$ Heterostructures Grown by Chemical Vapor Deposition}},\ }\bibfield  {journal} {\bibinfo  {journal} {ACS Applied Materials and Interfaces}\ }\textbf {\bibinfo {volume} {8}},\ \href {https://doi.org/10.1021/acsami.5b08295} {10.1021/acsami.5b08295} (\bibinfo {year} {2016})\BibitemShut
  {NoStop}%
\bibitem [{\citenamefont {Tan}\ \emph {et~al.}(2017)\citenamefont {Tan}, \citenamefont {Xu}, \citenamefont {Sheng}, \citenamefont {Lau}, \citenamefont {Fan}, \citenamefont {Chen}, \citenamefont {Tweedie}, \citenamefont {Wang}, \citenamefont {Zhou},\ and\ \citenamefont {Warner}}]{Tan2017c}%
  \BibitemOpen
  \bibfield  {author} {\bibinfo {author} {\bibfnamefont {H.}~\bibnamefont {Tan}}, \bibinfo {author} {\bibfnamefont {W.}~\bibnamefont {Xu}}, \bibinfo {author} {\bibfnamefont {Y.}~\bibnamefont {Sheng}}, \bibinfo {author} {\bibfnamefont {C.~S.}\ \bibnamefont {Lau}}, \bibinfo {author} {\bibfnamefont {Y.}~\bibnamefont {Fan}}, \bibinfo {author} {\bibfnamefont {Q.}~\bibnamefont {Chen}}, \bibinfo {author} {\bibfnamefont {M.}~\bibnamefont {Tweedie}}, \bibinfo {author} {\bibfnamefont {X.}~\bibnamefont {Wang}}, \bibinfo {author} {\bibfnamefont {Y.}~\bibnamefont {Zhou}},\ and\ \bibinfo {author} {\bibfnamefont {J.~H.}\ \bibnamefont {Warner}},\ }\bibfield  {title} {\bibinfo {title} {{Lateral Graphene‐Contacted Vertically Stacked WS 2 /MoS 2 Hybrid Photodetectors with Large Gain}},\ }\href@noop {} {\bibfield  {journal} {\bibinfo  {journal} {Advanced Materials}\ }\textbf {\bibinfo {volume} {29}},\ \bibinfo {pages} {1702917} (\bibinfo {year} {2017})}\BibitemShut {NoStop}%
\bibitem [{\citenamefont {Raja}\ \emph {et~al.}(2017)\citenamefont {Raja}, \citenamefont {Chaves}, \citenamefont {Yu}, \citenamefont {Arefe}, \citenamefont {Hill}, \citenamefont {Rigosi}, \citenamefont {Berkelbach}, \citenamefont {Nagler}, \citenamefont {Sch{\"u}ller}, \citenamefont {Korn} \emph {et~al.}}]{Raja2017}%
  \BibitemOpen
  \bibfield  {author} {\bibinfo {author} {\bibfnamefont {A.}~\bibnamefont {Raja}}, \bibinfo {author} {\bibfnamefont {A.}~\bibnamefont {Chaves}}, \bibinfo {author} {\bibfnamefont {J.}~\bibnamefont {Yu}}, \bibinfo {author} {\bibfnamefont {G.}~\bibnamefont {Arefe}}, \bibinfo {author} {\bibfnamefont {H.~M.}\ \bibnamefont {Hill}}, \bibinfo {author} {\bibfnamefont {A.~F.}\ \bibnamefont {Rigosi}}, \bibinfo {author} {\bibfnamefont {T.~C.}\ \bibnamefont {Berkelbach}}, \bibinfo {author} {\bibfnamefont {P.}~\bibnamefont {Nagler}}, \bibinfo {author} {\bibfnamefont {C.}~\bibnamefont {Sch{\"u}ller}}, \bibinfo {author} {\bibfnamefont {T.}~\bibnamefont {Korn}}, \emph {et~al.},\ }\bibfield  {title} {\bibinfo {title} {Coulomb engineering of the bandgap and excitons in two-dimensional materials},\ }\href@noop {} {\bibfield  {journal} {\bibinfo  {journal} {Nature Communications}\ }\textbf {\bibinfo {volume} {8}},\ \bibinfo {pages} {1} (\bibinfo {year} {2017})}\BibitemShut {NoStop}%
\bibitem [{\citenamefont {Sarkar}\ \emph {et~al.}(2014)\citenamefont {Sarkar}, \citenamefont {Liu}, \citenamefont {Xie}, \citenamefont {Anselmo}, \citenamefont {Mitragotri},\ and\ \citenamefont {Banerjee}}]{Sarkar2014}%
  \BibitemOpen
  \bibfield  {author} {\bibinfo {author} {\bibfnamefont {D.}~\bibnamefont {Sarkar}}, \bibinfo {author} {\bibfnamefont {W.}~\bibnamefont {Liu}}, \bibinfo {author} {\bibfnamefont {X.}~\bibnamefont {Xie}}, \bibinfo {author} {\bibfnamefont {A.~C.}\ \bibnamefont {Anselmo}}, \bibinfo {author} {\bibfnamefont {S.}~\bibnamefont {Mitragotri}},\ and\ \bibinfo {author} {\bibfnamefont {K.}~\bibnamefont {Banerjee}},\ }\bibfield  {title} {\bibinfo {title} {{MoS$_2$ field-effect transistor for next-generation label-free biosensors}},\ }\href@noop {} {\bibfield  {journal} {\bibinfo  {journal} {ACS Nano}\ }\textbf {\bibinfo {volume} {8}},\ \bibinfo {pages} {3992} (\bibinfo {year} {2014})}\BibitemShut {NoStop}%
\bibitem [{\citenamefont {Avsar}\ \emph {et~al.}(2020)\citenamefont {Avsar}, \citenamefont {Ochoa}, \citenamefont {Guinea}, \citenamefont {Zyilmaz}, \citenamefont {{Van Wees}},\ and\ \citenamefont {Vera-Marun}}]{Avsar2020}%
  \BibitemOpen
  \bibfield  {author} {\bibinfo {author} {\bibfnamefont {A.}~\bibnamefont {Avsar}}, \bibinfo {author} {\bibfnamefont {H.}~\bibnamefont {Ochoa}}, \bibinfo {author} {\bibfnamefont {F.}~\bibnamefont {Guinea}}, \bibinfo {author} {\bibfnamefont {B.}~\bibnamefont {Zyilmaz}}, \bibinfo {author} {\bibfnamefont {B.~J.}\ \bibnamefont {{Van Wees}}},\ and\ \bibinfo {author} {\bibfnamefont {I.~J.}\ \bibnamefont {Vera-Marun}},\ }\bibfield  {title} {\bibinfo {title} {{Colloquium: Spintronics in graphene and other two-dimensional materials}},\ }\href@noop {} {\bibfield  {journal} {\bibinfo  {journal} {Reviews of Modern Physics}\ }\textbf {\bibinfo {volume} {92}},\ \bibinfo {pages} {021003} (\bibinfo {year} {2020})}\BibitemShut {NoStop}%
\bibitem [{\citenamefont {Schaibley}\ \emph {et~al.}(2016)\citenamefont {Schaibley}, \citenamefont {Yu}, \citenamefont {Clark}, \citenamefont {Rivera}, \citenamefont {Ross}, \citenamefont {Seyler}, \citenamefont {Yao},\ and\ \citenamefont {Xu}}]{Schaibley2016}%
  \BibitemOpen
  \bibfield  {author} {\bibinfo {author} {\bibfnamefont {J.~R.}\ \bibnamefont {Schaibley}}, \bibinfo {author} {\bibfnamefont {H.}~\bibnamefont {Yu}}, \bibinfo {author} {\bibfnamefont {G.}~\bibnamefont {Clark}}, \bibinfo {author} {\bibfnamefont {P.}~\bibnamefont {Rivera}}, \bibinfo {author} {\bibfnamefont {J.~S.}\ \bibnamefont {Ross}}, \bibinfo {author} {\bibfnamefont {K.~L.}\ \bibnamefont {Seyler}}, \bibinfo {author} {\bibfnamefont {W.}~\bibnamefont {Yao}},\ and\ \bibinfo {author} {\bibfnamefont {X.}~\bibnamefont {Xu}},\ }\bibfield  {title} {\bibinfo {title} {{Valleytronics in 2D materials}},\ }\href@noop {} {\bibfield  {journal} {\bibinfo  {journal} {Nature Reviews Materials}\ }\textbf {\bibinfo {volume} {1}},\ \bibinfo {pages} {1} (\bibinfo {year} {2016})}\BibitemShut {NoStop}%
\bibitem [{\citenamefont {Wang}\ \emph {et~al.}(2015)\citenamefont {Wang}, \citenamefont {Tang}, \citenamefont {Sachs}, \citenamefont {Barlas},\ and\ \citenamefont {Shi}}]{Wang2015}%
  \BibitemOpen
  \bibfield  {author} {\bibinfo {author} {\bibfnamefont {Z.}~\bibnamefont {Wang}}, \bibinfo {author} {\bibfnamefont {C.}~\bibnamefont {Tang}}, \bibinfo {author} {\bibfnamefont {R.}~\bibnamefont {Sachs}}, \bibinfo {author} {\bibfnamefont {Y.}~\bibnamefont {Barlas}},\ and\ \bibinfo {author} {\bibfnamefont {J.}~\bibnamefont {Shi}},\ }\bibfield  {title} {\bibinfo {title} {Proximity-induced ferromagnetism in graphene revealed by the anomalous hall effect},\ }\href@noop {} {\bibfield  {journal} {\bibinfo  {journal} {Phys. Rev. Lett.}\ }\textbf {\bibinfo {volume} {114}},\ \bibinfo {pages} {016603} (\bibinfo {year} {2015})}\BibitemShut {NoStop}%
\bibitem [{\citenamefont {Wei}\ \emph {et~al.}(2016)\citenamefont {Wei}, \citenamefont {Lee}, \citenamefont {Lemaitre}, \citenamefont {Pinel}, \citenamefont {Cutaia}, \citenamefont {Cha}, \citenamefont {Katmis}, \citenamefont {Zhu}, \citenamefont {Heiman}, \citenamefont {Hone}, \citenamefont {Moodera},\ and\ \citenamefont {Chen}}]{Wei2016}%
  \BibitemOpen
  \bibfield  {author} {\bibinfo {author} {\bibfnamefont {P.}~\bibnamefont {Wei}}, \bibinfo {author} {\bibfnamefont {S.}~\bibnamefont {Lee}}, \bibinfo {author} {\bibfnamefont {F.}~\bibnamefont {Lemaitre}}, \bibinfo {author} {\bibfnamefont {L.}~\bibnamefont {Pinel}}, \bibinfo {author} {\bibfnamefont {D.}~\bibnamefont {Cutaia}}, \bibinfo {author} {\bibfnamefont {W.}~\bibnamefont {Cha}}, \bibinfo {author} {\bibfnamefont {F.}~\bibnamefont {Katmis}}, \bibinfo {author} {\bibfnamefont {Y.}~\bibnamefont {Zhu}}, \bibinfo {author} {\bibfnamefont {D.}~\bibnamefont {Heiman}}, \bibinfo {author} {\bibfnamefont {J.}~\bibnamefont {Hone}}, \bibinfo {author} {\bibfnamefont {J.~S.}\ \bibnamefont {Moodera}},\ and\ \bibinfo {author} {\bibfnamefont {C.~T.}\ \bibnamefont {Chen}},\ }\bibfield  {title} {\bibinfo {title} {Strong interfacial exchange field in the graphene/eus heterostructure},\ }\href@noop {} {\bibfield  {journal} {\bibinfo  {journal} {Nature Materials}\ }\textbf {\bibinfo {volume} {15}},\ \bibinfo {pages} {711}
  (\bibinfo {year} {2016})}\BibitemShut {NoStop}%
\bibitem [{\citenamefont {Mondal}\ \emph {et~al.}(2023)\citenamefont {Mondal}, \citenamefont {Biswas}, \citenamefont {Park}, \citenamefont {Cha}, \citenamefont {Kang}, \citenamefont {Yoon}, \citenamefont {Choi}, \citenamefont {Kim},\ and\ \citenamefont {Lee}}]{Mondal2023}%
  \BibitemOpen
  \bibfield  {author} {\bibinfo {author} {\bibfnamefont {A.}~\bibnamefont {Mondal}}, \bibinfo {author} {\bibfnamefont {C.}~\bibnamefont {Biswas}}, \bibinfo {author} {\bibfnamefont {S.}~\bibnamefont {Park}}, \bibinfo {author} {\bibfnamefont {W.}~\bibnamefont {Cha}}, \bibinfo {author} {\bibfnamefont {S.-H.}\ \bibnamefont {Kang}}, \bibinfo {author} {\bibfnamefont {M.}~\bibnamefont {Yoon}}, \bibinfo {author} {\bibfnamefont {S.~H.}\ \bibnamefont {Choi}}, \bibinfo {author} {\bibfnamefont {K.~K.}\ \bibnamefont {Kim}},\ and\ \bibinfo {author} {\bibfnamefont {Y.~H.}\ \bibnamefont {Lee}},\ }\bibfield  {title} {\bibinfo {title} {{Low Ohmic contact resistance and high on/off ratio in transition metal dichalcogenides field-effect transistors via residue-free transfer}},\ }\href@noop {} {\bibfield  {journal} {\bibinfo  {journal} {Nature Nanotechnology}\ } (\bibinfo {year} {2023})}\BibitemShut {NoStop}%
\bibitem [{\citenamefont {Mukherjee}\ \emph {et~al.}(2023)\citenamefont {Mukherjee}, \citenamefont {Dutta}, \citenamefont {Ghosh},\ and\ \citenamefont {Koren}}]{Mukherjee2023}%
  \BibitemOpen
  \bibfield  {author} {\bibinfo {author} {\bibfnamefont {S.}~\bibnamefont {Mukherjee}}, \bibinfo {author} {\bibfnamefont {D.}~\bibnamefont {Dutta}}, \bibinfo {author} {\bibfnamefont {A.}~\bibnamefont {Ghosh}},\ and\ \bibinfo {author} {\bibfnamefont {E.}~\bibnamefont {Koren}},\ }\bibfield  {title} {\bibinfo {title} {Graphene-in$_2$se$_3$ van der waals heterojunction neuristor for optical in-memory bimodal operation},\ }\href@noop {} {\bibfield  {journal} {\bibinfo  {journal} {ACS Nano}\ }\textbf {\bibinfo {volume} {17}},\ \bibinfo {pages} {22287} (\bibinfo {year} {2023})}\BibitemShut {NoStop}%
\end{thebibliography}%

\end{document}